\documentclass[aps,amsmath,amssymb,prd,nofootinbib,reprint,longbibliography]{revtex4-1}
\pdfoutput=1 % to force arxiv to pdflatex

\usepackage{graphicx}
\usepackage{hyperref}
\usepackage[utf8]{inputenc}

\newcommand{\comment}[1]{}
\newcommand{\beq}[1]{\begin{equation}\label{#1}}
\newcommand{\eeq}{\end{equation}}
\newcommand{\bea}{\begin{eqnarray}}
\newcommand{\eea}{\end{eqnarray}}
\renewcommand{\t}{\tilde}
\renewcommand{\b}{\bar}
\newcommand{\h}{\hat}
\newcommand{\del}{\partial}
\newcommand{\Del}{\nabla}

\newcommand{\M}{\ensuremath{\mathcal{M}}}
\newcommand{\N}{\ensuremath{\mathcal{N}}}
\newcommand{\w}{\wedge}
\newcommand{\G}{\ensuremath{\mathcal{G}}}
\newcommand{\F}{\ensuremath{\mathcal{F}}}
\newcommand{\Q}{\ensuremath{\mathcal{Q}}}
\renewcommand{\P}{\ensuremath{\mathcal{P}}}
\newcommand{\dbar}{{d\mkern-7mu\mathchar'26\mkern-2mu}}

\begin{document}

\title{Dirac branes for Dirichlet branes: Supergravity actions}
\thanks{Dedicated to the memory of J.~Polchinski}

\author{Andrew R.~Frey}\email{a.frey@uwinnipeg.ca}
\affiliation{Department of Physics and Winnipeg Institute for Theoretical
Physics\\ University of Winnipeg\\
515 Portage Avenue, Winnipeg, Manitoba R3B 2E9, Canada}

\date{\today}

\begin{abstract}Nontrivial Bianchi identities with local magnetic sources are 
solved by 
recognizing that gauge potentials are sections rather than globally defined
functions, but properly accounting for the source degrees of freedom requires
a modification of the field strength. Following work by Teitelboim and by
Cariglia and Lechner, we extend Dirac's string formalism for monopoles to
D-branes in type IIA and IIB string theory. We give novel derivations of 
brane-induced Chern-Simons terms in the supergravity actions, including a 
prescription for integrating over potentials in the presence of magnetic 
sources. We give a noncovariant formulation of the IIB theory, keeping only
the independent degrees of freedom of the self-dual 4-form potential. Finally,
it is well known that D8-branes source the mass parameter of IIA supergravity;
we show that the additional couplings of the massive IIA supergravity,
including on other D-brane worldvolumes, are a consequence of the corresponding
Dirac branes.\end{abstract}

%\keywords{D-branes, Supergravity Models, Solitons Monopoles and Instantons}

\maketitle

\section{Introduction}

In modern mathematical treatments, we recognize the vector potential of a
gauge theory not as a globally defined function but as the section of a 
gauge fiber bundle \cite{Wu:1976ge,Wu:1976qk}. In somewhat more 
pedestrian terms, the vector potential can be defined as different 
vector-valued functions in different coordinate patches of spacetime as long
as the distinct vector potentials are related by a gauge transformation on
the overlap of the coordinate patches (we will refer to this as ``gauge
patching'' of the vector potential). Gauge patching allows the description,
for example, of a constant magnetic field strength on a torus (the distinct
patches cover different unit cells) or the field of a magnetic monopole
(where the Bianchi identity $dF_2\neq 0$ can have no globally defined
solution). Analogs of both these examples for higher-rank form potentials
are important in string theory as harmonic background flux in compactifications
and higher-dimensional D-branes (and NS5-branes) that carry magnetic charges
for the fundamental potentials. The coupling between the magnetic current
and the potential is implicit in the patching and does not appear in the
action for the magnetic charge.

An alternative that displays the coupling of magnetic sources explicitly is to
double the number of gauge degrees of freedom by introducing dual field
strengths and corresponding potentials. In this so-called ``democratic'' 
formalism, the magnetic sources enter in the equations of motion (EOM) for the 
dual potentials \cite{hep-th/0103233}.\footnote{Including auxiliary fields
to enforce duality constraints, the IIA and IIB supergravities are given
in a democratic formalism in \cite{Bandos:2003et,DallAgata:1998ahf} 
respectively.} 
The extra degrees of freedom are then removed by enforcing duality
conditions $F_{D-p-2}=\pm\star F_{p+2}$ at the level of the EOM. In the 
democratic formalism, the action for magnetic charges includes the same 
current-potential coupling as for electric charges, so the EOM of the magnetic
charges includes the dual field strengths. Nontrivial Bianchi identities 
are enforced by the duality conditions. As a result, democratic formalisms 
still require gauge patching around magnetic sources.

Because the gauge transformations in the transition regions between gauge 
patches
are part of the definition of the potentials and also depend on the dynamical
magnetic currents, the gauge potentials are not independent degrees of 
freedom --- they have a hidden dependence on the magnetic brane degrees of
freedom which should be considered explicit in the language of calculus of 
variations. In quantum mechanical language, we need to separate the brane
and gauge degrees of freedom to serve as integration variables in the path
integral.

Interestingly, Dirac \cite{Dirac:1948um} provided a solution
in his early work on magnetic monopoles,\footnote{In fact, for monopoles,
\cite{Brandt:1977ks,Brandt:1976hk} showed that Dirac's formalism is 
equivalent to defining potentials as sections in part by showing that some 
gauge transformations move the Dirac string.}
which Teitelboim \cite{Teitelboim:1985yc}, Bandos \textit{et al.} 
\cite{Bandos:1997gd}, 
and Lechner and co-workers
\cite{hep-th/0103161,hep-th/0007076,hep-th/0107061,hep-th/0203238,hep-th/0302108,hep-th/0402078,hep-th/0406083}
extended to magnetic $p$-branes. 
The key idea is as follows:
Consider a field strength $\t F_{p+2}$ satisfying Bianchi identity
$d\t F_{p+2}=\beta\star j_{D-p-3}$ with $\beta$ a sign convention. Any conserved
current can be written as $\star j_{D-p-3}=d\star J_{D-p-2}$,\footnote{Assuming 
there 
are no harmonic forms on the full noncompact spacetime.} so a field strength 
defined as $\t F_{p+2}\equiv dC_{p+1}+\beta \star J_{D-p-2}$ satisfies the Bianchi 
identity; in this way, the magnetic coupling appears in the action. 
There is actually one additional subtlety when the branes fill
all noncompact dimensions; $\star j_{D-p-3}=d\star J_{D-p-2}$ on a compact 
manifold is inconsistent when there is net local charge, 
so we must define instead $\star j_{D-p-3}-\star j_{D-p-3}^*=d\star J_{D-p-2}$, 
where 
$j_{D-p-3}^*$ is some specified reference current (see \cite{arXiv:1807.07401}
for details in the case of magnetic monopoles). Now, $C_{p+1}$ is patched
around the reference current, so dynamics of $j_{D-p-3}$ do not affect the
potential. The alert reader may note that a Gauss law constraint means that 
the net charge must vanish on a compact manifold, but in string theory
charge may 
dissolve in background flux, so the net charge of local objects need not
vanish. Reference currents are necessary to account for this fact.

There are a variety of choices for the form $J$ for a given magnetically
charged $p$-brane with worldvolume $\M$ and current $j_\M$. As described by
\cite{Teitelboim:1985yc}, we can consider a ``Dirac $(p+1)$-brane'' with
worldvolume $\N$ of boundary $\del\N=\M-\M^*$. Then $J_\N$, the current of the
Dirac brane, satisfies $d\star J_\N=\star(j_\M-j_{\M^*})$. As we will see 
below, both the brane and Dirac brane currents are delta-function supported.
A less singular option for $J$, used by 
\cite{hep-th/0107061,hep-th/0203238,hep-th/0302108,hep-th/0402078,hep-th/0406083}, 
is given by the Chern kernel \cite{Chern1944,Chern1945}, 
which diverges only as a power law
near the current $j_\M$. In the following, we will mostly remain agnostic
about the nature of $J$, as our results are independent of this choice, but
we will often use the language of Dirac branes to be concrete and refer to 
$J$ as the Dirac brane current as shorthand. It is also
worth noting that, even fixing to Dirac branes or Chern kernels, $J$ is
arbitrary up to its co-derivative, but the field strength $\t F$ is invariant.

Our goal is to extend this formalism to D-branes in the IIA and
IIB string theories, writing these theories in terms of Ramond-Ramond
(RR) potentials $C_{p+1}$
for $p\leq 3$. While \cite{hep-th/0406083} have already considered
(arbitrary intersections of) D-branes in the IIB theory by means of an anomaly
argument, we present a new derivation via duality from the democratic 
formulation. In fact, \cite{hep-th/0406083} found several
new brane-induced terms in the IIB supergravity action, beyond the standard
coupling between currents and potentials, ie, the Wess-Zumino (WZ) action
for the D-branes. One set of new terms couples the Dirac brane
currents of magnetic branes to those of electric branes; 
\cite{Deser:1997se} first
identified the analogous term in Maxwell electrodynamics. These terms are
related to charge quantization. Further, \cite{hep-th/0406083} found
a correction to the bulk Chern-Simons (CS) term involving Dirac brane currents.
We will emphasize how these terms are required for consistency of the 
EOM and for gauge invariance. A key point in this story
is that D-brane currents are not conserved due to the WZ couplings,
but CS terms in the EOM and Bianchi identities cancel the anomaly via an
inflow argument \cite{hep-th/9512219,hep-th/0201221}; we give a detailed
accounting of the anomaly inflow in a general theory similar to that of
the RR forms. The necessity of reference currents also forces us to explain 
what it means to integrate over a gauge-patched potential.

The plan of this paper is as follows. In section \ref{s:inflow}, we 
demonstrate the anomaly inflow argument of \cite{hep-th/9512219,hep-th/0201221}
for a class of theories of form potentials which includes the RR potentials
of both type II supergravities. We pay particular attention to how the
inflow argument requires specific relations between various conventional
coefficients in the EOM and Bianchi identity and confirm the consistency
of the Dirac brane current with the inflow. Then
we consider an action principle for the generalized
theory of section \ref{s:inflow} and introduce the modified CS term in 
section \ref{s:actions} through a novel derivation. 
In section \ref{s:integration},
preliminary to our discussion of the supergravity actions, we find a 
prescription for integrating potentials that are gauge patched around 
magnetic sources, focusing on RR potentials in the 10D supergravities.
We then give a novel derivation of the new terms in the type IIB supergravity
action that were first described in \cite{hep-th/0406083} in section 
\ref{s:typeIIB}. We also eliminate
redundant degrees of freedom in the self-dual 4-form potential, leading to
a noncovariant action for the RR fields and discuss gauge invariance. 
Finally, in section \ref{s:typeIIA}, we derive the IIA supergravity action
including the Romans mass term \cite{Romans:1985tz} and D-branes for the 
first time. It has long been known that D8-branes
source the Romans mass, and we show for the first time 
how the corresponding Dirac 9-brane 
currents reproduce the additional couplings of the massive IIA supergravity.
We also propose that additional WZ couplings on D-branes in the 
massive theory \cite{hep-th/9603123,hep-th/9604119} 
are a consequence of the Dirac brane currents and conjecture the presence of 
other new WZ couplings on type IIA D-branes. We conclude with a brief 
discussion of future directions and give our conventions and some
auxiliary results in the appendices. 
A forthcoming companion paper \cite{freynew} 
will demonstrate how Dirac's formalism
separates the brane and gauge degrees of freedom in a form useful for 
dimensional reduction.

%%%%%%
\section{Currents and anomaly inflow}\label{s:inflow}

In this section, we describe general brane currents and when they are not
conserved. We then see how anomaly inflow determines the coefficients of 
several terms in the EOM and Bianchi identities and verify that the inflow
mechanism is always consistent with the Dirac brane formalism. Finally,
we apply our results to the 10D type II supergravity theories, making 
explicit the allowed, self-consistent sign conventions.

\subsection{Currents and anomalies}\label{s:currents}

Mathematically speaking, currents are dual vectors to differential forms
(see \cite{griffithsharris} for a review), which includes form integration 
over submanifolds of the appropriate dimensionality. In this respect, the
WZ action of a $p$-brane is the sum of $n$-currents acting on $n$-form
potentials:
\bea
S_{WZ}&=&\sum_{q=0}^{p+1}\Gamma_{\M_{p+1},\G_q^{(p)}}(C_{p+1-q})\label{Ecouplings}\\
&=&\mu_p \int_\M \left(\sum_{q=0}^{p+1} [C_{p+1-q}]\w\G_q^{(p)}\right)
=\mu_p\int_\M\, [C]\w\G^{(p)}, \nonumber\eea
where $\M_{p+1}$ is the worldvolume,
$\G_q$ are a series of worldvolume $q$-forms defined on the brane, and
$\mu_p$ is the $p$-brane charge. 
These may be defined to include pullbacks of spacetime forms.  
After the second equality, we have defined $C$ and $\G^{(p)}$ as formal 
sums over the 
various rank forms [we will suppress the superscript $(p)$ on $\G$ when the
dimensionality or type of brane is clear from context].

Of course, the WZ action (\ref{Ecouplings}) only describes a brane's electric
couplings to the gauge fields. This is sufficient in a democratic formulation
but potentials of all ranks do not exist when only independent degrees of 
freedom are included. As a result, a spacetime description of currents is
crucial. Since any dual vector $\Gamma_n$ is uniquely identified with a 
differential form $j_n$ by the inner product
\beq{innerprod} \Gamma_{\M,\G}(C_n)\equiv \int C_n\w \star j_{\M,\G} \eeq
with the integration over spacetime, we can identify the $j_{\M,\G}$ as the
brane currents. Then the EOM and Bianchi identity 
\bea
d\star\t F_{p+2}&=&(-1)^{D-p-1}\star j_{p+1}+\cdots\quad\text{and}\nonumber\\
d\t F_{p+2} &=& \beta_{p}\star j_{D-p-3}+\cdots\label{eombianchi} \eea
give the electric and magnetic couplings, where $\beta_{p}$ is a sign chosen 
by convention.\footnote{The sign on the current in the EOM is determined by
the canonical action 
\bea
S&=& \int d^Dx\sqrt{-g}\sum_p\left[-\frac{1}{2} |\t F_{p+2}|^2+
C_{p+1}\cdot j_{p+1}\right]\label{action1}\\
&=&\int \sum_p(-1)^{p(D-p)+1}\left[\frac 12 \t F_{p+2}\w\star\t F_{p+2}
+(-1)^D C_{p+1}\w \star j_{p+1}\right] .\nonumber\eea}
The current $j_{p+1}=\sum j_{\M_{p+q+1},\G_q}$ with the sum over all branes.
(On the flip side, all the currents of the same brane can be written as a
formal sum $j_\M=\sum_q j_{\M,\G_q}$.)

Naively, the current for a $(p+q)$-brane with worldvolume form $\G_q$ is
\beq{branecurrents}
j^{\mu_1\cdots\mu_{p+1}}_{\M,\G}(x) = \mu_{p+q}\int_{\M_{p+q=1}}\h dX^{\mu_1}\w\cdots 
\h dX^{\mu_{p+1}}\w\G_q\,\delta^D(x,X) ,
\eeq
where $X^\mu$ are the embedding coordinates. This may be modified in 
topologically nontrivial situations, such as when the brane in question is
actually the nontransversal intersection of two other branes. Nontransversal
intersections are the focus of 
\cite{hep-th/9710206,hep-th/0402078,hep-th/0406083}. 
As our goal is to emphasize writing the action in terms of independent
degrees of freedom, we base our results on the naive current 
(\ref{branecurrents}); the adaptation to nontransversal intersections follows
from \cite{hep-th/0406083}. 

The anomalies we consider are local in nature, so they must cancel
pointwise.  While related, we emphasize that these anomalies are 
separate from global anomalies that forbid certain brane configurations,
such as the Freed-Witten anomaly \cite{hep-th/9907189} or the magnetic D-brane
Gauss law constraint that $H_3$ integrate to zero over the worldvolume
(see \cite{hep-th/0201029,hep-th/0505208}). From the perspective of the WZ
action (\ref{Ecouplings}), they arise from a gauge variation 
$\delta C=d\lambda$. Integrating the pullback by parts yields a term from the
brane boundary and one from $\h d\G$. In some cases, these can cancel
between branes; for example, a D1-brane can end on a D3-brane, providing 
a magnetic source for the D3-brane gauge field. The two anomalous terms cancel
in the summed current $j_2$.\footnote{In fact, this configuration can also
be described as a BIon solution of the D3-brane theory, in which case
there is manifestly no anomaly.} 
However, if $\h d\G$ contains the pullback of a
spacetime form, the cancellation must be by inflow associated with a 
modified gauge transformation $\delta C$ as occurs in string theory.

From a spacetime point of view, we can consider the divergence of the brane
current
\begin{widetext}\bea
(\star d\star j_{\M,\G})^{\mu_1\cdots\mu_{p}}&=&(-1)^{(p+1)(D-p)}\mu_{p+q}
\int_\M \Del_\nu
\left[\delta^D(x,X)\h d X^\nu\w\h dX^{\mu_1}\w\cdots \h dX^{\mu_p}\w\G_q\right]
\nonumber\\
&=&(-1)^{D(p+1)+1}\mu_{p+q}\int_\M \h d\left[\delta^D(x,X)\h dX^{\mu_1}\w \cdots 
\h dX^{\mu_{p}}\w \G_q\right]\nonumber\\
&&+(-1)^{D(p+1)+p}\mu_{p+q}\int_\M \h d X^{\mu_1}\w\cdots\h dX^{\mu_p}\w\h d\G_q\,
\delta^D(x,X)\ .
\label{divcurrent}\eea\end{widetext}
In an arbitrary Lorentzian metric, $\delta^D(x,X)$ is the biscalar 
distribution, and the covariant derivative acts with respect to the 
spacetime position $x$, but the derivative switches to the partial with
respect to $X$ as described in \cite{arXiv:1102.0529}.
We therefore see that the brane currents are not conserved:
\beq{nonconservation}
d\star j_{\M,\G}=(-1)^{D+q}\star\t\jmath_{\del\M,\G}-(-1)^{D}\star j_{\M,\h d\G},\eeq
where $\t\jmath$ is a $p$-form current for the boundary of the worldvolume.
Henceforth in this paper, we will assume that the boundary contributions
cancel with some of the worldvolume $\h d\G$ contributions, so we will ignore
those terms from here out, returning to them in the companion paper 
\cite{freynew}.

Consider then a single brane (ie, $\del\M=0$) with $\h d\G$ a 
pullback of a nontrivial spacetime form $H_{r+1}$.
We will find that the anomalies cancel 
when $\h d\G_q = \sum_r \eta_{p,q,r} [H_{r+1}]\G_{q-r}$, where $H_{r+1}$ is 
prototypically the field strength of a potential that does not
couple to the branes, and $\eta_{p,q,r}$ is some proportionality constant.
Then, using (\ref{idents}),
\bea
\star j_{\M,[H_{r+1}]\G_{q-r}} &=& \star\left(j_{\M,\G_{q-r}}\cdot H_{r+1}\right)
\label{pullback2wedge}\\
&=&(-1)^{(r+1)(D-r-1)}H_{r+1}\w\star j_{\M,\G_{q-r}} .\nonumber
\eea
We will find that $\eta$ is independent of the rank $q$ of the worldvolume 
forms, so the anomaly for the total brane current is conveniently written as
\beq{nonconservation2}
d\star j_{p+1}= \sum_r (-1)^{r(D-r)}\eta_{p,r}H_{r+1}\w\star j_{p+r+1} .\eeq

\subsection{Anomaly inflow}\label{s:eombianchiinflow}

We consider a set of potentials $C_{p+1}$ (the RR potentials in string theory)
and corresponding gauge-invariant field strengths $\t F_{p+2}$, with an 
additional set of field strengths 
$H_{r+1}=dB_r$ assumed closed with pullbacks $[H_{r+1}]$ that appear in 
$\h d\G$ for some branes (with this coupling, those branes carry an electric
current for $B_r$, but $H_{r+1}$ remains closed).  
The classical anomaly discussed in the previous section then 
appears in the current whenever $H_{r+1}\neq 0$, whether it is a 
topologically nontrivial flux or due to another brane source. It is a simple
generalization to add an extra index to $C$ or $B$ to have more than one
potential at each rank. Our discussion of the inflow is similar to comments
by \cite{hep-th/9512219} for M-theory and is implicit in \cite{hep-th/0406083} 
for IIB string theory; \cite{hep-th/0201221} 
gives a worldvolume argument for string theory. We are not 
aware of a discussion in this full class of theories.

The general EOM for $C_{p+1}$ (to first order in $\t F_{p+2}$) has the structure
\bea
d\star\t F_{p+2}&=&(-1)^{D-p-1}\star j_{p+1}\nonumber\\
&&+\sum_{r=0}^{D-p-3}\left[\alpha_{p,r} (\star\t F_{p+r+2})\w H_{r+1}\right.
\nonumber\\
&&\left.+ \t\alpha_{p,r}\t F_{D-p-r-2}\w H_{r+1}\right] ,\label{eomF2}\eea
where $\alpha,\t\alpha$ are constants.  Meanwhile, the Bianchi identity is
\beq{bianchiF2}
d\t F_{p+2} = \beta_p \star j_{D-p-3}+\sum_{r=0}^{p+1}\t\beta_{p,r}\t F_{p-r+2}\w
H_{r+1} ;\eeq
$\beta_p$ is a sign convention, which can be chosen independently for
each field strength.  For now, we treat the $\alpha,\t\alpha,\beta,\t\beta$
as independent constants, though there are relations among them in a 
Lagrangian formulation of the theory; other conditions following from gauge
invariance are discussed in appendix \ref{a:invariance}. 
The $\t\alpha$ terms follow from 
CS terms in the action, while the $\alpha$ and $\t\beta$ terms arise from 
terms in the field strength. To distinguish them from CS terms, we will
refer to the $\alpha$ and $\t\beta$ terms as ``transgression'' terms.
Theories with this structure include of course the type II supergravities 
and also the dimensionally reduced theory of gravity and form potentials
on a torus, for example. 

Current (non)conservation is related to the integrability conditions obtained
by taking the exterior derivative of the EOM and Bianchi identity.
For the EOM, we have
\bea
d\star j_{p+1}&=& \sum_r \left[-(-1)^r \alpha_{p,r}+(-1)^{D-p}\t\alpha_{p,r}
\beta_{D-p-r-4}\right]\nonumber\\
&&\times\star j_{p+r+1}\w H_{r+1},\label{eomintegrability}\eea
leaving off terms that are independent of the brane currents.\footnote{These
must vanish separately. We discuss how this occurs and the relation to
gauge invariance in appendix \ref{a:invariance}.}
In other words, we see that the derivatives of the CS and transgression terms 
localize on the currents as needed for an anomaly inflow.
Comparing to equation (\ref{nonconservation2}), we find
\beq{eominflow}
\eta_{p,r}=-(-1)^{D+(p+1)(r+1)}\alpha_{p,r}-(-1)^{pr}\t\alpha_{p,r}\beta_{D-p-r-4}\eeq
since the anomaly must cancel when only one $H_{r+1}$ background is nonvanishing.
In fact, this holds for any case where only one brane contributes to the
current, so we see that $\eta$ is independent of the worldvolume form rank.

Similarly, the Bianchi identity yields
\beq{bianchiintegrability}
d\star j_{D-p-3} = -\beta_p \sum_r \t\beta_{p,r}\beta_{p-r}\star j_{D-p+r-3}\w H_{r+1}
.\eeq
Cancellation of the anomaly then requires
\beq{bianchiinflow}
\eta_{D-p-4,r}=(-1)^{r(D-p)+p}\beta_p\beta_{p-r}\t\beta_{p,r} .
\eeq 
Again, we see that $\eta$ is independent of $q$.

\subsection{Dirac brane currents}\label{s:diracbranes}

We will now see how Dirac brane currents in the field strengths fit into
the Bianchi identities using the constraint (\ref{bianchiinflow}). 
Our discussion extends similar results in \cite{hep-th/0406083}.

To start, we need to consider how to extend the worldvolume form \G\ from the
brane worldvolume to the Dirac brane or, in the case $J$ represents a Chern
kernel, the entire spacetime. The key point is to choose the reference brane
worldvolume $\M^*$ homotopic to \M\ (so \N\ is continuous for a Dirac brane).
The extension of \G\ depends on its form. For D-branes, we will be 
concerned primarily with the case that $\G=\G([B_2],\h dA_1)$, where $A_1$
is the worldvolume gauge field. In this case, we continue to take $B_2$ as 
given by the spacetime NSNS form (pulled back to \N\ as appropriate). 
For the gauge field, we choose a fixed $A_1^*$ on $\M^*$ and an extension 
$\b A_1$ to \N\ or spacetime that pulls back to $A_1$ on \M\ and $A_1^*$ on 
$\M^*$. Then the Dirac brane current takes the form of (\ref{branecurrents}) 
with $\G$ promoted to the extension $\b\G$ and $X^\mu$ replaced by the 
embedding coordinates $Y^\mu$ of \N. For a Chern kernel of a $(p+q)$-brane,
$\star J_{\b\G}\equiv (-1)^{(p+1)(q+1)}(\star J)\b\G$ (a similar formula holds
for Dirac brane currents when $\b\G$ is the pullback of a spacetime form
by virtue of (\ref{idents})).

Since the Dirac brane contribution to the field strength is given by the
Dirac brane's current, equation (\ref{nonconservation}) applies in the form
\beq{nonconservation3}
d\star J_{\N,\b\G} =(-1)^{D+q}\star\left(j_{\M,\G}-j_{\M^*,\G}\right)-
(-1)^D\star J_{\N,\b d\b\G} .\eeq
Therefore, to cancel the dynamical current $j_{D-p-3}=\sum j_{\M_{D-p+q-1},\G_q}$
(summed over all branes) in the Bianchi identity (\ref{bianchiF2}), 
we should define
\beq{transgression}
\t F_{p+2}=dC_{p+1}+(-1)^D\beta_p\star J_{D-p-2}
+\sum_{r=0}^{p+1} \t\beta_{p,r}C_{p-r+1}\w H_{r+1} ,\eeq
where the Dirac brane current is a sum over the corresponding Dirac branes
\beq{diracbranecurrent}
J_{D-p-2} = \sum (-1)^q J_{\N_{D-p+q-2},\b\G_q} \eeq
and $C_{p+1}$ includes the potential for the reference current for worldvolumes
$\M^*$. Then the current for the Dirac brane associated with a given physical
brane can be written as a formal sum $J_\N= \sum_q (-1)^q J_{\N,\b\G_q}$.

With this definition for the total Dirac brane current, the divergence
(\ref{nonconservation3}) and condition (\ref{bianchiinflow}) give
\begin{widetext}\bea
d\star J_{D-p-2}&=&(-1)^D\star(j_{D-p-3}-j^*_{D-p-3}) -\beta_p\sum_{q,r}
(-1)^{q+(r+1)(p-r-1)}\t\beta_{p,r}\beta_{p-r}H_{r+1}\w\star J_{\N,\b\G_{q-r}}\nonumber\\
&=&(-1)^D\star(j_{D-p-3}-j^*_{D-p-3})+\beta_p\sum_r\t\beta_{p,r}\beta_{p-r}
\star J_{D-p+r-2}\w H_{r+1} .
\label{diracdiv}
\eea
Since $C_{p+1}$ contains the potential for the reference current, 
\bea
d\t F_{p+2} &=& \beta_p\star j_{D-p-3}+\sum_r \t\beta_{p,r}\left(dC_{p-r+1}+\sum_\ell
\t\beta_{p-r,\ell}C_{p-r-\ell+1}\w H_{\ell+1}+\beta_{p-r}\star J_{D-p+r-2}\right)
\w H_{r+1}\nonumber\\
&=& \beta_p\star j_{D-p-3}+\sum_r \t\beta_{p,r}\t F_{p-r+2}\w H_{r+1} .
\label{bianchireduced}\eea\end{widetext}
In other words, the co-derivative of the Dirac brane current is precisely
consistent with the appearance of Dirac brane currents in the transgression
term.

\subsection{Type II supergravity conventions}\label{s:sugraconventions}

We can now apply our results to set limits on possible sign conventions
in the 10D type II supergravities. Some of the restrictions we find below
(such as the alternating of signs in the duality conditions for the
democratic formulation) appear implicitly in the literature (see for example
the discussion of conventions in appendix A of \cite{arXiv:1609.00385}), 
but we are not
aware of an explicit derivation from first principles.

While the D-brane WZ action identifies 
\beq{DbraneG}
\G=e^{\F}\w 
\sqrt{\frac{\hat A(4\pi^2\alpha' R_T)}{\hat A(4\pi^2\alpha' R_N)}} ,
\eeq
where $ \F=2\pi\alpha' F_2+\eta[B_2]$, $\h A$ is the A-roof genus, and $R_T,R_N$
are the tangent and normal bundle curvatures, we will not consider the
$\alpha'$ corrections, instead restricting to $\G=\exp\F$. See 
\cite{hep-th/0406083} for more on $\alpha'$ corrections in the IIB theory.
Both type II supergravities in this approximation have a single background
field strength $H_3=dB_2$ and, following from the above, a common sign
choice $\eta=\eta_{p,2}$ appearing in $\F$ for all $p$. Here, we will also
follow the typical choice of setting all $\t\beta_{p,2}\equiv\t\beta$, a 
single sign choice for the transgression terms in each theory.

Starting with the IIB theory with only potentials $C_{p+1}$ for $p\leq 3$, 
the EOM and Bianchi identities are
\bea
d\star\t F_1&=&\star j_0+\alpha_{-1}\star\t F_3\w H_3,\nonumber\\
 d\t F_1&=&\beta_{-1}\star j_8,\nonumber\\
d\star\t F_3&=&\star j_2+(\alpha_1\star\t F_5+\t\alpha_1\t F_5)\w H_3,\nonumber\\
d\t F_3&=&\beta_1\star j_6+\t\beta \t F_1\w H_3,\nonumber\\
d\star\t F_5&=&\star j_4+\t\alpha_3\t F_3\w H_3 ,\nonumber\\
d\t F_5&=&\beta_3\star j_4+\t\beta \t F_3\w H_3 .\label{IIBeqns}\eea
The appearance of both transgression and CS terms for the $\t F_3$ EOM is due
to the self-duality condition on $\t F_5$.
The constraint (\ref{eominflow}) tells us immediately that
$\eta=-\alpha_{-1}=-\t\alpha_3\beta_1=-\alpha_1-\t\alpha_1\beta_3$. 
Meanwhile, the Bianchi
identities are consistent with (\ref{bianchiinflow}) for
$\eta=-\beta_{-1}\beta_1\t\beta =-\beta_{1}\beta_3\t\beta$. Finally, the
transgression terms in the EOM and Bianchi identity are related through 
variation of the Lagrangian, leading to equation (\ref{coeffs}), which 
implies $\alpha_{-1}=-\t\beta$ and $\eta=\t\beta$. 
All told, there are two independent
sign choices, $\t\beta$ and $\beta_3$, with the signs in the Bianchi 
identities alternating $\beta_3=-\beta_1=\beta_{-1}$. 

The type IIA supergravity (including a possible mass term) has
\bea
d\star\t F_0&=&0,\quad d\t F_{0}=\beta_{-2}\star j_9\nonumber\\
d\star\t F_2 &=&-\star j_1+\alpha_0\star\t F_4\w H_3 ,\quad d\t F_2= \beta_0
\star j_7+\t\beta\t F_0\w H_3\nonumber\\
d\star \t F_4&=&\star j_4+\t\alpha_2\t F_4\w H_3,\quad d\t F_4=
\beta_2\star j_5+\beta\t F_2\w H_3 .\label{IIAeqns}\eea
As in the IIB case, we find $\eta=\alpha_0=-\t\alpha_2\beta_2$ and 
$\eta\t\beta=\beta_{-2}\beta_0=\beta_0\beta_2$. Derivation of the transgression
terms from the action gives also $\alpha_0=-\t\beta$, which tells us that
$\eta=-\t\beta$ and $\beta_0=-\beta_{-2}=-\beta_2$. There are once again
two independent sign choices, with the others determined.

In either supergravity, the democratic formulation has
\bea
d\star\t F_{p+2} &=&(-1)^{p+1}\star j_{p+1}+\alpha_p \star\t F_{p+4}\w H_3 ,
\nonumber\\
d\t F_{p+2}&=&\beta_p\star j_{7-p}+\t\beta \t F_p\w H_3\label{democratic}\eea
for $-2\leq p\leq 8$ ($\t F_{p<0}\equiv 0$). 
By comparison to the Bianchi identities above, 
the duality relations must be $\star\t F_{D-p-2}=\mp \beta_p\t F_{p+2}$ (in IIA
and IIB respectively) for $p\leq 3$; in particular, $C_4$ satisfies the
self-duality relation $\star\t F_5=\beta_3\t F_5$. 
Since the coefficients $\beta_p$ alternate signs in each
theory, so do the duality relations. Since the D-brane charge (vs antibrane)
is determined by the WZ coupling to $C_{p+1}$ in the democratic formulation,
these alternating signs mean that D$p$-branes with $p\geq 3$ enter the
Bianchi identities with alternating signs as well. The signs can only be
chosen the same if the transgression coefficients $\t\beta_p$ are distinct
for the different $\t F_{p+2}$. Finally, since the Bianchi identities for the
higher-rank field strengths have the same form, we have $\beta_p\beta_{p+2}=-1$
(i.e., alternating signs) for all $p$.

%%%%%%%%
\section{Brane-modified Chern-Simons action}\label{s:actions}

Chern-Simons terms are familiar from the actions of both 10D type II 
supergravities and the 11D supergravity. We emphasize here that the presence
of D-branes necessarily modifies those CS terms; by extension (through
duality, etc), M-branes in 11D and NS5-branes in 10D must also modify them. 
The CS term modifications were first pointed out by \cite{hep-th/0406083};
here we give a new, simple, physically-motivated derivation in the theory
of the previous section, which we can apply to the 10D supergravities later.

The action 
\bea
S&=&\frac{1}{2\kappa_0^2}
\int \sum_p(-1)^{p(D-p)+1}\left[\frac 12 \t F_{p+2}\w\star\t F_{p+2}\right.
\nonumber\\
&&\left. +(-1)^D C_{p+1}\w \star j_{p+1}\vphantom{\frac 12}
\right]+S_{CS},\label{action}\eea
with Chern-Simons terms
\bea
S_{CS} &=& \frac{1}{2\kappa_0^2}\sum_{p,r}\int \left[\gamma_{p,r}C_{p+1}\w 
\t F_{D-p-r-2}\right.\nonumber\\
&&\left.+\t\gamma_{p,r}(\star J_{D-p-2})\w C_{D-p-r-3}\vphantom{\t F}
\right]\w H_{r+1}\label{CSaction}
\eea
reproduces the $C_{p+1}$ EOM (\ref{eomF2}), assuming that the 
field strength is defined by equation (\ref{transgression}) with the Dirac
brane current (\ref{diracbranecurrent}). $\gamma_{p,r},\t\gamma_{p,r}$ are 
some set of constants related to the EOM coefficients $\t\alpha_{p,r}$.
We have ignored kinetic terms for the closed field strengths $H_{r+1}$ as well
as the gravitational sector.
The canonical coupling between the potential and electric current in 
(\ref{action}) determines the sign of the source term in (\ref{eomF2}), and
it also gives the WZ action for all the branes (i.e., there is no need for
an additional WZ action on the branes) up to a factor of the gravitational
coupling $2\kappa_0^2$, which can be accommodated by rescaling the brane
charges.

We discuss the invariance of this action and the field strength 
(\ref{transgression}) under the gauge transformations of $C_{p+1}$ in the
absence of brane sources
in appendix \ref{a:invariance}, arriving at constraints
(\ref{gaugeconstraint1},\ref{gaugeconstraint4}), which also guarantee
that the Bianchi identity and EOM depend only on $\t F_{p+2}$ rather than
$C_{p+1}$ (in the absence of currents). 
We also find the relationships (\ref{coeffs}) between the 
EOM coefficients $\alpha_{p,r},\t\alpha_{p,r}$ and $\t\beta_{p,r},\gamma_{p,r}$
in the appendix. 

While the $\gamma_{p,r}$ terms in $S_{CS}$ are familiar from, for example, the 10D
supergravities, the $\t\gamma_{p,r}$ terms require some explanation.
Without them, the Dirac brane currents are absent from the field strengths 
in the $\t\alpha_{p,r}$ terms of the EOM. That would leave the EOM dependent
on the Dirac brane worldvolumes \N\ including the arbitrary reference branes
$\M^*$, which would clearly be inconsistent (in fact, that would be a 
violation of gauge invariance in the magnetic context). So we consider
\bea
\frac{\delta S_{CS}}{\delta C_{p+1}}&=& \sum_r \gamma_{p,r}\t F_{D-p-r-2}\w H_{r+1}
\label{vary1}\\
&&+\sum_r (-1)^{p(D-p-r)}\gamma_{D-p-r-4,r}\left(\t F_{D-p-r-2}\right.\nonumber\\
&&\left.-(-1)^D\beta_{D-p-r-4}\star J_{p+r+2}\right.\nonumber\\
&&\left.\vphantom{\t F} +(-1)^{(p+1)(D-p-r)}\t\gamma_{D-p-r-4,r}
(\star J_{p+r+4})\right)\w H_{r+1}\nonumber\eea
assuming the coefficients obey (\ref{gaugeconstraint4}). Therefore, we 
require $\t\gamma_{p,r}=(-1)^{D-p}\beta_p\gamma_{p,r}$ to ensure that the EOM
are written only in terms of the field strengths.

We will later derive these and other new brane-induced terms for the type II
supergravities.

%%%%%%%%
\section{Integrating patched potentials}\label{s:integration}

It is not immediately clear what it means to integrate over a quantity
including a potential with gauge patching because the potential is not
single valued in the overlap of coordinate patches: Either gauge is physically
acceptable. As a result, many authors, including 
\cite{hep-th/9605033,hep-th/9710206,hep-th/0406083}
have suggested writing potential-current
couplings in terms of the gauge-invariant field strength. However, if we 
attempt to write an action following that approach without the explicit 
appearance of the potentials, the EOM will contain the arbitrary reference
currents $j^*_{p+1}$ (for electric sources). 
Consider, for example, the action (\ref{action}) above
with the replacement $C_{p+1}(\star j_{p+1})\to(-1)^p\t F_{p+1}(\star J_{p+2})$. 
Even ignoring transgression terms by setting $H_{r+1}\to 0$, the variation
of this term is $\delta C_{p+1}\w \star (j_{p+1}-j_{p+1}^*)$.
Here we give what is to our knowledge the first description of how to carry 
out spacetime integrals including gauge-patched potentials.

The key idea is already present in \cite{Wu:1976qk}, who gave a prescription 
for integrating the vector potential of Maxwell theory along a charged
particle worldline in the presence of a monopole. In sketch form our new
prescription for integration against other forms over spacetime is as
follows: Pick an
arbitrary division where the integrated potential switches from one gauge
to another. Then design the integral to be invariant under changes of 
the division, a choice closely related to gauge invariance. 
Since we are integrating over spacetime, we also have to 
exclude the locus of magnetic charge, where the potential is undefined.

For simplicity, we work with RR potentials in string theory. Specifically,
we consider a set of potentials $C_{p+1}$ for $p$ either even or odd (as in
the IIA or IIB supergravity respectively). 
Written as a formal sum of forms, the gauge-invariant field strengths are
$F=dC+\t\beta C H_3$ and the gauge transformations are 
$\delta C=d\chi-\t\beta \chi H_3$, where $\t\beta$ is a single sign choice. 
Depending on the application, the formal sum $C$ could include $p=-1$ to $8$
as in a democratic formulation of the supergravity or only a subset (e.g.,
$p=0,2$ in type IIA supergravity). Taking $p=0,H_3=0$, our prescription
also applies to standard electrodynamics. The field strengths do not
include Dirac branes, and the Bianchi identities are 
$dF=\star (\beta j^*)+\t\beta F H_3$, where 
$(\beta j^*)=\sum_p\beta_p j_{p+1}^*$ with $\beta_p$ a distinct sign choice
for each potential. Our results apply to any gauge-patched potential,
meaning $j^*$ could represent dynamical currents, but we will take $j^*$ to
be fixed reference currents. It is worth recalling that a high-dimension brane
with a worldvolume gauge field or in the presence of nonvanishing $B_2$ 
contributes to lower-rank currents, so $j^*_{p+1}$ may include currents smeared
over worldvolumes with dimension greater than $p+1$.

To define the spacetime integral $\int CK$, where $K$ is another formal sum of
forms, we note that the potentials $C$ are undefined on the collection of
branes that contribute to $j^*$. Contrast this to electric charges where
potentials simply diverge; for the example of a magnetic monopole, none of
the coordinate patches with well-defined potentials covers the monopole 
location. Therefore, rather than integrating over all of spacetime, we excise
a small tube around the worldvolume of each magnetically charged brane with
boundary \P. After integrating, we will take the volume of the excluded tube
to vanish, so it becomes measure zero. In the presence of multiple magnetic
brane sources, \P\ has multiple components.

\begin{figure}[t]
\centering
\includegraphics[width=0.5\textwidth]{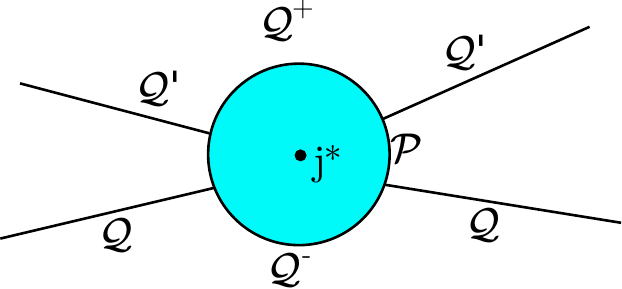}
\caption{Sketch of integration region with magnetic current $j^*$. Shaded
region inside \P\ is excluded; \Q,$\Q'$ are possible separations of two gauge
patches.\label{f:integration}}
\end{figure}

Now consider the overlap region of two gauge patches (multiple gauge patches 
are a straightforward extension, assuming the configuration is simple enough)
of potentials $C^\pm$ around some magnetic source. 
On the overlap region, the two potentials are related
by the gauge transformation $C^+=C^-+d\zeta-\t\beta \zeta H_3$.
We choose a codimension-one surface \Q\ in the overlap region
such that the spacetime volume $\Q^\pm$ on either side of \Q\ is within
the coordinate patch where $C^\pm$ is valid respectively. This is just a 
partition of spacetime into regions where each potential is used.
Note that \Q\ will generically intersect the boundary surface \P. 
In the example of a monopole, this is just related to the fact that the
region where each potential is valid is given by a range of polar angle.
Figure \ref{f:integration} sketches the various surfaces for a current $j^*$.
Quantities integrated over \Q\ or \P\ are assumed to be pulled back to the
appropriate submanifold.

We now define the spacetime integral as
\beq{integral1}
\int C\w K \equiv \int_{\Q^+} C^+\w K + \int_{\Q^-} C^-\w K \pm\int_\Q \zeta\w K ,
\eeq
where the sign on the \Q\ integral depends on the orientation of the 
integration measure. We will choose the positive sign below and take care to
account for signs due to the orientation.

This is sensible provided that the integral is unchanged under changes of
the arbitrary dividing surface \Q. Consider then changing $\Q\to\Q'$.
The change in the integral is 
\bea
&&\int_\Q^{\Q'} (C^- - C^+)\w K +\int_{\Q'}\zeta\w K -\int_\Q \zeta\w K \nonumber\\
&&=\int_\Q^{\Q'}\zeta\w\left[(-1)^pdK +\t\beta H_3\w K\right] -\int_{\b\P}
\zeta\w K ,\label{integral2} \eea
where $\int_\Q^{\Q'}$ indicates the region with boundary $\Q'-\Q-\b\P$
(that is, the region between \Q\ and $\Q'$ in figure \ref{f:integration})
and $\b\P$ is the region on \P\ between its intersections with $\Q,\Q'$.
This will vanish provided $dK+(-1)^p\t\beta H_3 K=0$\footnote{Note that
$p$ is either even or odd for all potentials, so the same condition holds
for all terms in the formal sum.} and either $K$ has no
delta-function-like singularity (so the $\b\P$ integral vanishes as \P\ 
shrinks) or $\zeta K=0$ (for example, due to the legs of each form). 
It is not a coincidence that $dK+(-1)^p\t\beta H_3 K=0$ is the same condition
for the integral (\ref{integral1}) to be gauge invariant under gauge 
transformations $\chi$ that are globally defined over the integration region
and vanish on the boundary at infinity and \P. 

Maxwell electrodynamics provides some simple examples. Consider a static
monopole of charge $g$ at the origin. 
Then the simplest form for \P\ is a sphere of radius
$\epsilon$ around the origin. With the typical choices 
$A_1^\pm= g(\pm 1-\cos\theta)d\phi$, \Q\ is any surface that intersects
\P\ in the limit $\epsilon\to 0$ and does not intersect the $z$-axis.
A simple choice for $Q$ is the $xy$-plane with transition function
$\zeta=2g\phi$. This prescription for integration
also works for harmonic flux with no magnetic sources on a compact manifold.
For example, we can consider a constant magnetic field on a square $T^2$, 
which we can describe by vector potential $A_1=By dx$ in the first unit cell
$0\leq x,y<2\pi R$. To make the potential periodic in $y$, we must work
in a different gauge in each unit cell given by $2\pi R n\leq y<2\pi R(n+1)$
with gauge transition function $\zeta=-2\pi RBx$, but of course each gauge
is valid over the entire covering space. To integrate over the first
unit cell, we choose any curve \Q\ that runs from $x=0$ to $x=2\pi R$ within
the unit cell. A simple choice for \Q\ is the $x$-axis.

Integration by parts requires some care but is sensible for formal sums
$K=dk+(-1)^p\t\beta H_3 k$. We start by integrating by parts in $\Q^\pm$
separately to find
\bea
\int C\w K &=& (-1)^p\int F\w k\label{byparts1}\\ &&+\int_\Q \left[\zeta\w K
+(-1)^p(C^+-C^-)\w k\right]\nonumber\\
&&-(-1)^p\int_{\P^+}C^+\w k-(-1)^p\int_{\P^-}C^-\w k ,\nonumber\eea
where $\P^\pm=\P \cap \Q^\pm$. Note that the integrals over $\P^\pm$ are
signed based on the orientation of the integration measure. The integral
over \Q\ is simply a surface term $-(-1)^p\zeta k$ over \P, which combines
with the $\P^\pm$ integrals. We have
\beq{byparts2}
\int C\w K = (-1)^p\left[\int F\w k-\int_\P C\w k\right] .\eeq
The first integral on the right-hand side
is over the same region as the original integral, i.e., all spacetime exterior
to \P, but it can extend to all spacetime since $F$ is globally defined
(assuming $k$ is globally defined).

It is tempting to think that the \P\ integral in equation (\ref{byparts2})
vanishes as \P\ shrinks. However, because $C$ does not have a well-defined
limit at the location of $j^*$, that is not always the case. There are three
cases of special interest. First, suppose that $k=\star J$ is given by a 
Dirac brane current with boundary at other locations (so $J$ ``passes through''
\P). As a given component of \P\ shrinks, the pullback of $C$ approaches
the potential of the magnetic source in \P, which reverses orientation 
compared to the $\star J$ on either side of \P. 
As a second case, suppose $k=\star J$ is the Dirac
brane current emanating from the brane source inside \P. Then $J$ is aligned
along the worldvolume $\M^*$ and the radial direction inside \P, so $\star J$
has the same components as $C$ on \P\ as \P\ shrinks, and the integral
again vanishes. Finally, suppose that $k=d\kappa+(-1)^p\t\beta H_3\kappa$.
Since \P\ has no boundary, integration by parts gives
\beq{byparts3}
(-1)^p\int_\P C\w k = \int_\P F\w \kappa .\eeq
Assuming $\kappa$ is sufficiently smooth inside \P, we can replace the latter
by the integral of $(dF)\kappa$ over the excluded region inside \P\ in the 
limit as \P\ shrinks. Then, we can replace $dF$ using the Bianchi identity
and keep only the delta-function-like term $\beta\star j^*$. After careful
accounting of signs,
\beq{byparts4}
\int C\w K = (-1)^p\int F\w k+\int (\beta\star j^*)\w\kappa ,\eeq
where both integrals on the right-hand side are taken over all spacetime.

In the above, we have ignored possible complications from brane intersections
or overlaps of three or more gauge patches in a nontrivial configuration. 
Taking care with these will 
potentially reproduce the modifications \cite{hep-th/0406083} 
noted are necessary for nontransversal intersections.

Finally, we note that a variation of the potential $C$ can include a 
variation of the gauge transition function $\zeta$, but it need not. In 
particular, if $C^\pm \to C^\pm +\delta C$ for a globally defined variation
$\delta C$, the variation of the integral is well defined over all of
spacetime. No special prescription is needed for the integration.

%%%%%%
\section{Type IIB supergravity}\label{s:typeIIB}

We now turn to the main result of this paper, the type II supergravity
actions in the presence of D-branes, starting with the IIB case. 
We start by reminding the reader of the bosonic IIB action in the absence
of D-branes and using the result of section \ref{s:actions} to find the 
modification to the CS term. Then we provide a novel derivation of this
term and other new terms involving Dirac brane currents by dualizing the
democratic formulation of the IIB supergravity. Keeping 10D covariance and
a self-dual 5-form field strength, we find agreement with
\cite{hep-th/0406083}, modulo terms involving anomalies on brane intersections
and $\alpha'$ corrections, which we do not consider. We also give a new
analysis of gauge invariance for this action. Finally, we separate
the 4-form potential and 5-form field strength into electric and magnetic
components and write a noncovariant action with D-brane contributions
in terms of the independent degrees of freedom only.

\subsection{Action and modified CS terms}

The gauge-invariant field strengths of the type IIB supergravity are
\bea
\t F_1 &=& dC_0+\beta_{-1}\star J_9 ,\quad \t F_3=dC_2+\t\beta C_0\, H_3
+\beta_1\star J_7 ,\nonumber\\
\t F_5&=& dC_4+\t\beta C_2\w H_3+\beta_3\star J_5 ,\label{fieldIIB}
\eea
using the convention that all transgression terms have the same sign $\t\beta$.
Recall that $\beta_3=-\beta_1=\beta_{-1}$. Excluding local sources, the
action of the bosonic sector is
\bea
S_{IIB}&=&\frac{1}{2\kappa_0^2}\int d^{10}x\sqrt{-g}e^{-2\phi}\left[R+4(\del\phi)^2
\right]\label{actionIIB}\\
&&+\frac{1}{2\kappa_0^2}\int\left[\frac 12 e^{-2\phi}H_3\w\star H_3
+\frac 12 \t F_1\w\star \t F_1\right.\nonumber\\
&&\left.+\frac 12\t F_3\w\star\t F_3
+\frac 14\t F_5\w\star\t F_5+\frac 12\beta_3\t\beta\, 
C_4\w\t F_3\w H_3\right] .\nonumber
\eea
The coefficient of the CS term is determined by the results of section
\ref{s:sugraconventions} and equation (\ref{coeffs}). Note that the $\t F_5$
kinetic term is halved because it is self-dual and contains duplicate degrees
of freedom.

As in section \ref{s:actions}, we can add D-brane currents to (\ref{actionIIB})
by adding Dirac brane currents to the field strengths and shifting the action by
\bea
S_{IIB}&\to& S_{IIB}+\frac{1}{2\kappa_0^2}\int\left[C_0\w\star j_0+
C_2\w\star j_2\right.\nonumber\\
&&\left.+\frac 12 C_4\w\star j_4-\frac 12 \t\beta
(\star J_5)\w C_2\w H_3\right] .\label{actionIIBshift}
\eea
Here, $j_0$ is a scalar current associated with D$(-1)$-instantons, and 
the last term, the CS term modification, includes the Dirac 4-brane current 
as required to make the EOM gauge invariant. The coupling to $C_4$ has a 
factor of $1/2$ due to the 5-form self-duality. 
We also note that the relationship (\ref{coeffs}) between coefficients of the
action and EOM implies that we can replace 
\bea
&&\frac{\t\beta}{2}\int\left[\beta_3 C_4\w\t F_3\w H_3-(\star J_5)\w C_2\w H_3
\right]\nonumber\\
&&\ \rightarrow -\frac{\t\beta}{2} \int\left[\beta_3 C_2\w\t F_5\w H_3+
(\star J_7)\w C_4\w H_3\right]\label{CSreplace}
\eea
in the action to generate the same EOM for the RR gauge fields. (We will see
below that there are actually other terms also.)
Integration by parts to make this
replacement (up to surface terms) follows along the lines of section 
\ref{s:integration} with a slight modification to account for the fact that
both potentials $C_2,C_4$ are patched. The integral on the boundary 
\P\ vanishes, and the replacement (\ref{CSreplace}) holds with both
forms of the CS term following the prescription of section \ref{s:integration}
for integration.

Finally, we note that the action is sometimes written with an additional
CS term (see for example \cite{Johnson:2003gi})
\beq{extraCSIIB}
\frac{1}{2\kappa_0^2}\int\frac 14 B_2\w C_2 \w dB_2\w dC_2 .\eeq
In the absence of branes, this term is a total derivative and does not
contribute to the EOM. Even in the presence of D5-branes, it can be 
written as the integral of $d(C_2^2 d(B_2)^2)$ following the integration
prescription given above. However, with both D5- and NS5-branes, this 
term seems to be nonvanishing. Understanding its completion in the 
presence of all sources and whether it remains topological is a task we
leave to the future.

\subsection{Dualization from democratic formulation}\label{s:dualization}

The EOM given in (\ref{democratic}) for the RR potentials 
of type IIB supergravity in the democratic 
formulation are given by the (pseudo)action
\bea
S_{IIB,dem}&=&\frac{1}{2\kappa_0^2} \int \left[\frac 14 \t F_1\w\star\t F_1
+\frac 14 \t F_3\w\star\t F_3 +\frac 14 \t F_5\w\star\t F_5\right.\nonumber\\
&&\left. +
\frac 14 \t F_7\w\star\t F_7+\frac 14 \t F_9\w\star\t F_9
+\frac 12 C_0\w\star j_0  +\frac 12 C_2\w\star j_2\right.\nonumber\\
&&\left.
+\frac 12 C_4\w\star j_4+\frac 12 C_6\w\star j_6 +\frac 12 C_8\w\star j_8\right]
\label{democraticIIB}
\eea
(dropping the Einstein-Hilbert and dilaton and $B_2$ kinetic terms for 
convenience) with the field strengths given by equation (\ref{fieldIIB})
and
\beq{fieldIIB2}
\t F_7=dC_6+\t\beta C_4\w H_3+\beta_5\star J_3 ,\quad
\t F_9=dC_8+\t\beta C_6\w H_3+\beta_7\star J_1 .\eeq
These also reproduce the Bianchi identities in 
(\ref{democratic}).\footnote{Note that we are not considering type I 
supergravity, so we do not include a D9-brane current, but it is a 
straightforward generalization since $C_{10}$ does not have a dual potential.}
The EOM must be supplemented by duality relations between the higher- and
lower-rank field strengths. If we instead enforce the definitions 
(\ref{fieldIIB2}) with Lagrange multipliers and identify those Lagrange
multipliers with the lower-rank field strengths, we can generate an
equivalent action for only the lower-rank potentials.
As in section \ref{s:integration}, the terms with potentials are over 
spacetime with the reference currents removed, but other terms can be 
integrated over the entire spacetime since they are smooth at the reference
currents and the punctures are zero measure.

Now we turn to removing the extra degrees of freedom from the action
(\ref{democraticIIB}). Start by adding Lagrange multipliers
\bea
S_{IIB,dem}&\to& S_{IIB,dem}+\frac{1}{2\kappa_0^2}\int\frac 12\left[\lambda_1
\w\left(\t F_9-dC_8\right.\right. \nonumber\\
&&\left.\left.-\t\beta C_6\w H_3-\beta_7\star J_1\right)\right.
\label{lagrangeIIB}\\
&&\left.+\lambda_3\w \left(\t F_7-dC_6-\t\beta C_4\w H_3-\beta_5\star J_3\right)
\right] \nonumber
\eea
and treat $\t F_{7,9}$ as independent degrees of freedom.
The $\t F_{7,9}$ EOM give $\lambda_1=\star\t F_9,\lambda_3=\star\t F_7$.
Meanwhile, varying $C_{6,8}$ gives EOM
\beq{EOMpotIIB}
d\lambda_1=\star j_8,\quad d\lambda_3=\star j_6+\t\beta H_3\w\lambda_1 .\eeq
These are consistent with $\lambda_1\equiv\beta_{-1}\t F_1$ and 
$\lambda_3\equiv\beta_1\t F_3$ recalling that $\beta_{-1}=-\beta_1$. 
Imposing these identifications is equivalent to imposing the duality relations
of the democratic formulation.

The action is linear in $C_{6,8}$, so it may seem that imposing the EOM
(\ref{EOMpotIIB}) eliminates them. However, there are nontrivial surface
terms of the form
\bea
-\frac{\beta_3}{2}&\!\!\!\!\!\!&\!\!\!\!\!\!
\int_\P \left(C_8\w\t F_1-C_6\w\t F_3\right)\nonumber\\
 &=& 
\frac{\beta_3}{2}\int\left(\beta_7 C_0\w\star j_0^*-\beta_5 C_2\w\star j_2^*
\vphantom{\frac 12}\right)\nonumber\\
&=&\frac 12\int\left(\vphantom{\frac 12}
C_0\w\star j_0^*+C_2\w\star j_2^*\right)\label{surfaceIIBduality} \eea
from integrating (\ref{lagrangeIIB}) by parts. [In the language of section
\ref{s:integration}, we have taken $C=C_4+C_6+C_8$ and 
$K=d\lambda_1+(d\lambda_3-\t\beta H_3\lambda_1)-\t\beta H_3\lambda_3$, so
\P\ surrounds $j^*_{0,2,4}$.] Note that the electric reference currents $j_{0,2}^*$
are not excised from the integrals over $C_{0,2}$ (though excised 
higher-dimensional reference branes may carry these currents).
At this point, the action has become
\begin{widetext}
\bea
S_{IIB}&=&\frac{1}{2\kappa_0^2} \int \left[\frac 12 \t F_1\w\star\t F_1
+\frac 12 \t F_3\w\star\t F_3 +\frac 14 \t F_5\w\star\t F_5 +
\frac 12 C_0\w \star(j_0+j_0^*)+\frac 12 C_2\w\star( j_2+j_2^*)\right.\nonumber\\
&&\left. +\frac 12 C_4\w \star j_4 +\frac 12 \beta_3\t\beta C_4\w\t F_3\w H_3
-\frac 12 \t F_1\w\star J_1-\frac 12 \t F_3\w\star J_3\right] .
\eea

Finally, the last two terms split into terms involving the potentials and
those involving only Dirac brane currents. The former integrate by parts
using equation (\ref{diracdiv}) in IIB supergravity form
\beq{diracdivIIB}
d\star J_{p+2} = \star (j_{p+1}-j_{p+1}^*)-\t\beta \star J_{p+4}\w H_3 \eeq
(surface terms on \P\ vanish by reasoning given in section \ref{s:integration}),
leaving current terms and a remainder involving $\star J_5$.
All together, we have
\bea
S_{IIB}&=&\frac{1}{2\kappa_0^2}\int \left[\frac 12 \t F_1\w\star \t F_1
+\frac 12\t F_3\w\star\t F_3 +\frac 14\t F_5\w\star\t F_5+C_0\w\star j_0+
C_2\w\star j_2+\frac 12 C_4\w\star j_4\right.\nonumber\\
&&\left. +\frac 12\beta_3\t\beta\, C_4\w\t F_3\w H_3-
\frac 12 \t\beta C_2\w(\star J_5)\w H_3+\frac 12\beta_3 \star J_1\w\star J_9
-\frac 12\beta_3\star J_3\w \star J_7\right] .\quad\qquad\label{actionIIB2}
\eea\end{widetext}

Some comments on the action (\ref{actionIIB2}) are in order. First, we note
the factor of $1/2$ on the $C_4-j_4$ coupling. It is well known that the
D3-brane charge must be reduced by half for the gauge EOM to work out correctly
for the self-dual 5-form (see \cite{hep-th/0105097} for example). 
In fact, we made that
choice for the same reason in equation (\ref{actionIIBshift}). Here we see
that it is a consequence of dualization from the democratic action, where
all the potential-current couplings are halved. Second, we note
the appearance of the usual CS term and also the Dirac brane modification,
both of which are a consequence of transgression terms in the field strength
and in the divergence of the Dirac brane current. Finally, there are two 
new terms involving Dirac brane currents for electrically and magnetically
charged branes. These terms were also found by \cite{hep-th/0406083} for IIB
supergravity and by \cite{Deser:1997se} for monopoles in electrodynamics.
In the monopole case, \cite{Deser:1997se} showed that these terms are 
topological (do not contribute to the classical EOM) but are related to
charge quantization. However, in the type II supergravities, the Dirac brane
currents depend not just on the brane positions but also the brane gauge
fields and $B_2$, so they may not be purely topological (in either IIA or IIB
supergravity). 
In our discussion of the IIA supergravity, we will see
that these terms can have physical importance even in the classical theory;
specifically, they will reproduce one of the CS terms of the massive IIA 
theory. Further discussion of the contribution of $(\star J)^2$-type terms to
classical EOM will appear in \cite{freynew}.

\subsection{Gauge invariance}

We have now provided two novel derivations of the modified CS term (and also
found $J^2$-type terms when starting with the democratic action).
As it has not been discussed previously in the literature, it is important
to understand invariance of $S_{IIB}$ from (\ref{actionIIB2}) under gauge
transformations of the RR potentials, particularly because the CS terms are 
not invariant on their own in the presence of branes, in contrast to the
usual presentation in the absence of branes. 
The gauge transformations take the form
$\delta C=d\chi-\t\beta \chi H_3$ for globally defined forms $\chi_p$. Since
the integrals including the potentials $C$ have boundary at infinity and \P,
the $\chi_p$ should vanish on \P.

Before considering the final action (\ref{actionIIB2}), 
it is worth commenting on the gauge invariance of (\ref{democraticIIB}). 
The variation of the action is
\bea
\delta S &=& \frac{1}{2\kappa_0^2} \int \frac 12 \left[ \chi_1\w\left(d\star j_2
-\t\beta H_3\w \star j_4\right)\right.\nonumber\\
&&\left.  + \chi_3\w\left(d\star j_4
-\t\beta H_3\w \star j_6\right)\right.\label{gaugeIIBdem}\\
&&\left. + \chi_5\w\left(d\star j_6
-\t\beta H_3\w \star j_8\right)+\chi_7\w d\star j_8\right] .\nonumber
\eea
This vanishes by virtue of equation (\ref{nonconservation2}) with the
identification $\eta=\t\beta$.

On the other hand, the variation of the potential-current couplings in
(\ref{actionIIB2}) cannot all cancel. Fortunately, in the presence of
branes, the CS terms are also not gauge-invariant on their own. The variation
of the action is
\bea
\delta S &=& \frac{1}{2\kappa_0^2} \int \left[ \chi_1\w d\star j_2+\frac 12
\chi_3\w d\star j_4-\frac 12 \t\beta \chi_1\w H_3\w\star j_4\right.\nonumber\\
&&\left.+\frac 12
\beta_3\t\beta \chi_3\w d\t F_3\w H_3-\frac 12 \t\beta \chi_1\w d\star J_5\w
H_3\right]\nonumber\\
&=& \frac{1}{2\kappa_0^2} \int \left[\t\beta \chi_1\w\star j_4\w H_3
\left(1-\frac 12 -\frac 12\right)\right.\nonumber\\
&&\left. +\t\beta \chi_3\w\star j_6\w H_3
\left(\frac 12-\frac 12\right)\right] =0 .\label{gaugeIIB}
\eea
The reference current $\star j_4^*$ does not appear because it lies in 
the removed punctures, and 
surface terms on \P\ vanish because the $\chi_p$ vanish on \P.
Note that the CS term with the Dirac brane current is necessary for invariance
under transformations of $C_2$.

\subsection{Noncovariant formulation for independent degrees of freedom}
\label{s:noncovariant}

To complete the action, \cite{hep-th/0406083} used 
auxiliary variables with the Pasti-Sorokin-Tonin (PST) 
formalism \cite{hep-th/9611100,hep-th/9701037}
to enforce the self-duality condition and maintain 10D 
covariance.\footnote{Sen \cite{arXiv:1511.08220,arXiv:1903.12196}
has developed an alternate formalism also using
auxiliary fields to describe self-dual field strengths.} 
Here, we determine the action for independent degrees of freedom only,
breaking 10D covariance and some of the $C_4$ gauge invariance. Making the
choice of degrees of freedom correctly can be useful in determining the
effective theory of a dimensional reduction; lack of 10D covariance is not
necessarily a disadvantage.

First, we need to identify independent degrees of freedom in $\t F_5$, which
we do by separating its components into two sets, ``electric'' components
$\t F_5^{(1)}$, which we treat as independent, and ``magnetic'' components 
$\t F_5^{(2)}$ that satisfy $\t F_5^{(2)}=\beta_3\star\t F_5^{(1)}$. 
The next task is to divide the potential also into
electric and magnetic components $C_4^{(1,2)}$. In some cases, it is possible
to make a clean division such that $C_4^{(1,2)}$ contribute only to 
$\t F_5^{(1,2)}$ respectively. This is true, for example, of the nonvanishing
components of $C_4$ for the K\"ahler moduli of Calabi-Yau compactifications
even in the presence of background flux and warping 
\cite{arXiv:0810.5768,arXiv:1308.0323,arXiv:1609.05904}. However, it is
not true in general. What is possible is to choose a set of magnetic components 
$C_4^{(2)}$ that does not appear in $\t F_5^{(1)}$, while the complementary
set of electric components $C_4^{(1)}$ appears in both $\t F_5^{(1,2)}$. If we
ignore gauge invariance, the numbers of components in these sets are different.

To proceed, we fix our spacetime coordinates, including a spatial coordinate
$\t x$ (with $g_{\t x\mu}=0$ for $\mu\neq\t x$ for simplicity; we consider only
the case where we can do so). Then we label
any form with a leg along $\t x$ with $(2)$ and any form without a leg on $\t x$
with $(1)$. A prototype coordinate for $\t x$ is a spatial direction in a 
Minkowski factor of a (possibly warped) product metric. This choice for
$\t F_5^{(1,2)}$ follows \cite{hep-th/0105097}, for example. For notational
convenience, we define $\t d=d\t x \del_{\t x}$ and $\dbar =d-\t d$. Then
\bea
\t F_5^{(1)}&=&\dbar C_4^{(1)}+\t\beta (C_2\w H_3)^{(1)}+\beta_3\star J_5^{(2)} ,
\nonumber\\
\t F_5^{(2)}&=&\dbar C_4^{(2)}+\t d C_4^{(1)}+
\t\beta (C_2\w H_3)^{(2)}+\beta_3\star J_5^{(1)} .\label{noncovF5}
\eea

To find the noncovariant action, we start with action (\ref{actionIIB2}) 
and project $\t F_5,(C_2H_3),J_5$ and $C_4$ onto electric and magnetic 
components as above. Adding in a Lagrange multiplier, the relevant part
of the action is
\bea
S&=&\frac{1}{2\kappa_0^2} \int\left[\cdots + \frac 14 \t F_5^{(1)}\w\star
\t F_5^{(1)}+\frac 14 \t F_5^{(2)}\w\star\t F_5^{(2)}\right.\nonumber\\
&&\left.
+\frac 12 C_4^{(1)}\w\star j_4^{(1)}+\frac 12 C_4^{(2)}\w\star j_4^{(2)}\right.
+\beta_3\t\beta C_4^{(1)}\w(\t F_3\w H_3)\nonumber\\
&&\left.
+\beta_3\t\beta C_4^{(2)}\w(\t F_3\w H_3)-\frac 12\t\beta\star J_5^{(1)}\w
(C_2\w H_3)\right.\nonumber\\
&&\left.-\frac 12\t\beta\star J_5^{(2)}\w(C_2\w H_3)
+\frac 12 \lambda_5\w\left(\t F_5^{(2)}-\dbar C_4^{(2)}-\t d C_4^{(1)}
\right.\right.\nonumber\\
&&\left.\left.-\t\beta(C_2\w H_3)^{(2)}
-\beta_3\star J_5^{(1)}\right)\right] .\label{selfdual1}
\eea
Note that a wedge product $A^{(1,2)} B$ chooses the $(2,1)$ components of
form $B$.  We can now follow the same procedure as in the previous subsection, 
finding $\lambda_5=\star\t F_5^{(2)}$ from the $\t F_5^{(2)}$ EOM, and the
duality relation $\lambda_5=\beta_3\t F_5^{(1)}$ plus $C_4^{(2)}$ EOM yield
\beq{noncovBianchi}
\dbar \t F_5^{(1)}=\t\beta (\t F_3\w H_3)^{(1)}+\beta_3\star j_4^{(2)} ,\eeq
which is the Bianchi identity following from (\ref{noncovF5}). It is also 
the part of the covariant Bianchi identity with no legs along $\t x$.
There are two remaining finer points in the derivation. First, 
we assume that $C_4^{(1)}$ is not patched on the surface surrounding $j_4^{(1),*}$
even though it appears in $\t F_5^{(2)}$. Second, we note that the 
projection of the relation (\ref{diracdivIIB}) onto the magnetic components
involves both Dirac brane currents:
\beq{noncovdiracdiv}
\dbar\star J_5^{(1)}+\t d\star J_5^{(2)} =\star (j_4^{(1)}-j_4^{(1),*})
-\t\beta\left(\star J_7\w H_3\right)^{(2)} .\eeq
In the end, we find
\begin{widetext}\bea
S&=&\frac{1}{2\kappa_0^2}\int\left[\cdots +\frac 12 \t F_5^{(1)}\w\star\t F_5^{(1)}
+C_4^{(1)}\w\star j_4^{(1)}+\frac 12\beta_3\t\beta C_4^{(1)}\w(\t F_3\w H_3)
-\frac 12\beta_3\t\beta \t F_5^{(1)}\w(C_2\w H_3)-\frac 12\beta_3\t F_5^{(1)}\w
\t d C_4^{(1)}\right.\nonumber\\
&&\left. -\frac 12\t\beta\star J_5^{(2)}\w(C_2\w H_3)
-\frac 12\t\beta C_4^{(1)}\w(\star J_7\w H_3)-\frac 12 C_4^{(1)}\w \t d \star J_5^{(2)}
+\frac 12 \beta_3\star J_5^{(1)}\w
\star J_5^{(2)}\right] .\label{selfdual2}
\eea\end{widetext}
Interestingly, the noncovariant action seems to mix
the two forms for the CS term equated in (\ref{CSreplace}). There are two
entirely new terms. 

To validate the action (\ref{selfdual2}), we can check that it gives the
correct equations of motion for the potentials. The $C_4^{(1)}$ EOM follows
after some substitution; to write the EOM in terms of field strengths only,
we must rewrite $\dbar\t d C_4^{(1)}$ in terms of $\t F_5^{(1)}$ and recall
that $(\t F_3H_3)^{(2)}=\dbar[(C_2H_3)^{(2)}]+\t d[(C_2H_3)^{(1)}]$.
After some cancellation,
\beq{C4EOMnoncov}
\dbar \star\t F_5^{(1)}+\beta_3\t d\t F_5^{(1)}=\star j_4^{(1)} 
+\beta_3\t\beta (\t F_3\w H_3)^{(2)} .\eeq
Using the duality relation, this is also the part of the covariant EOM with
one leg along $\t x$, as expected.
The $C_2$ EOM is slightly more subtle, as we must
move the projection from $\delta C_2 H_3$ to other factors in wedge products
to get the full variation. In particular, the second
CS term contains $(\delta C_2 H_3)^{(1)} (C_2H_3)=-\delta C_2 (C_2H_3)^{(2)} H_3$.
To combine this with other terms to make the gauge-invariant $\t F_5^{(1)}$,
we must notice that $0=C_2 H_3^2=(C_2H_3)^{(1)}H_3+(C_2H_3)^{(2)}H_3$.
Further, we must notice that the variation of 
$\t\beta C_4^{(1)} \t F_3 H_3-\t F_5^{(1)}\t d C_4^{(1)}$ yields 
$\t\beta(d-\t d)C_4^{(1)}H_3=\t\beta\dbar C_4^{(1)}H_3$, also required to write
the EOM in terms of $\t F_5^{(1)}$. We see that
\beq{C2EOMnoncov}
d\star\t F_3=\star j_2-\t\beta\star\t F_5^{(1)}\w H_3-\beta_3\t\beta\t F_5^{(1)}
\w H_3\ .\eeq
Again, that is exactly as expected from the decomposition of the usual $C_2$
EOM.

As an alternate derivation, the action (\ref{selfdual2}) is in principle 
equivalent to the action in section 5.3 of \cite{hep-th/0406083} with
the auxiliary scalar of the PST formalism gauge fixed to a specific form.
The remaining PST gauge symmetries (discussed for IIB supergravity in
\cite{DallAgata:1997gnw}) eliminate the components $C_4^{(2)}$.\footnote{We
thank D.~Sorokin for this and other provocative comments regarding this
subsection.} The companion paper \cite{freynew} will emphasize applications
where it is important to consider only independent degrees of freedom,
and (\ref{selfdual2}) will play a role there. This derivation highlights the
origin of the new terms in the covariant formulation.

%%%%%%
\section{Type IIA supergravity}\label{s:typeIIA}

In this section, we give the type IIA supergravity action with D-brane
sources for the first time. We will first derive the action from the 
democratic formulation, in which the brane-current couplings, i.e., the 
brane WZ terms, are known, following the same techniques as used for the IIB
theory. We address gauge invariance under the RR gauge transformations and
verify that the action we obtain is consistent with the constraints discussed
in section \ref{s:actions}. We then discuss the well-known relation between
D8-branes and the massive IIA supergravity in light of our new action.
We will examine the role that Dirac brane currents play in generating the
couplings of the massive IIA theory, including terms proportional to the
mass parameter in D-brane WZ actions.

\subsection{Action, gauge invariance, and EOM}

As for the IIB theory, we start with a democratic (pseudo)action
(for the RR sector)
\bea
S_{IIA,dem} &=& -\frac{1}{2\kappa_0^2} \int\left[ \frac 14 \t F_0\w\star\t F_0
+\frac 14 \t F_2\w\star\t F_2+\frac 14 \t F_4\w\star\t F_4\right.\nonumber\\
&&\left.+\frac 14 \t F_6\w\star\t F_6+\frac 14 \t F_8\w\star\t F_8 
+\frac 14 \t F_{10}\w\star\t F_{10}\right.\nonumber\\
&&\left. +\frac 12 C_1\w\star j_1+\frac 12 C_3\w\star j_3
+\frac 12 C_5\w\star j_5+\frac 12 C_7\w\star j_7\right.\nonumber\\
&&\left.+\frac 12 C_9\w\star j_9\right]\label{democraticIIA}
\eea
and field strengths defined by 
\bea
\t F_0&=&m+\beta_{-2}\star J_{10} ,\nonumber\\
\t F_2 &=& dC_1+\t\beta m B_2+\beta_0\star J_8 ,\nonumber\\
\t F_4 &=& dC_3+\t\beta C_1\w H_3+\frac{m}{2}B_2^2+\beta_2\star J_6 ,\nonumber\\
\t F_6 &=& dC_5+\t\beta C_3\w H_3+\frac{m}{3!}\t\beta B_2^3+
\beta_4\star J_4\nonumber\\
\t F_8 &=&dC_7 +\t\beta C_5\w H_3+\frac{m}{4!} B_2^4+\beta_6\star J_2\ ,
\nonumber\\
\t F_{10}&=&dC_9 +\t\beta C_7\w H_3+\frac{m}{5!}\t\beta B_2^5
+\beta_8\star J_0 .\label{IIAfields}
\eea
Some of the field strength definitions (\ref{IIAfields}) require an explanation.
First, consistent with the possible presence of D8-branes, we include
the $\t F_0$ and $\t F_{10}$ field strengths, and we include a ``bare'' mass
parameter $m$ obeying $dm=\beta_{-2}\star j_9^*$. This and the choice 
to add $m\exp(\t\beta B_2)$ to $\t F$ ensure that the Bianchi identities
of (\ref{IIAeqns}) are satisfied. These choices are consistent with the 
massive IIA supergravity \cite{Romans:1985tz}. 
Second, although $\t F_{10}$ automatically
has a trivial Bianchi identity simply by index counting, we include a 
transgression term consistent with the gauge transformation of $C_9$
that leaves the D8-brane WZ action invariant. We also include a 0-rank
Dirac brane current though there is not a D$(-2)$-brane or associated 
$j_{-1}$ current. Instead, we recall that each D$p$-brane has a series of
currents $j_{\M},j_{\M,\F},j_{\M,\F^2/2},\cdots$ and Dirac brane currents
$J_{\N},J_{\N,\b\F},J_{\N,\b\F^2/2},\cdots$; the rank-0 Dirac brane current from 
this series contributes to $J_0$, even though there is not a corresponding
D-brane current. To our knowledge, this is the first appearance of this current
in the literature.

From this point, derivation of the action for the lower-rank potentials
follows the same steps as in section \ref{s:dualization}. We find
\begin{widetext}\bea
S_{IIA}&=&-\frac{1}{2\kappa_0^2} \int\left[ \frac 12 \t F_0\w\star\t F_0
+\frac 12 \t F_2\w\star\t F_2+\frac 12 \t F_4\w\star\t F_4
+C_1\w\star j_1+ C_3\w\star j_3+\frac 12 \beta_0\t\beta C_3\w\t F_4\w H_3
\right.\nonumber\\
&&\left. +\frac m4 \beta_0\t\beta C_3\w B_2^2\w H_3
-\frac 12 \t\beta C_3\w\star J_6\w H_3
-\frac 12 \left(\frac{m}{4!}B_2^4-\beta_0\star J_2\right)\w\left(m\t\beta\beta_0
B_2+\star J_8\right)\right.\nonumber\\
&&\left.
+\frac 12\left(\frac{m}{3!}\t\beta B_2^3+\beta_0\star J_4\right)\w\left(
\frac m2 \beta_0 B_2^2-\star J_6\right)
+\frac 12 \t F_0\w\left(\frac{m}{5!}\t\beta\beta_0 B_2^5+\star J_0\right)
\right] .\label{actionIIA}
\eea\end{widetext}
As in the IIB theory, we find a modified CS term and couplings between 
Dirac brane currents of electrically charged and magnetically charged D-branes.
The last term is also of this type, but $\t F_0$ could also in principle 
include the mass parameter. 

It is worth noting that we recover the action of the pure massive IIA 
supergravity (i.e. with no D-branes) for $j_{p+1},J_{p+2}\to 0$ 
\cite{Romans:1985tz} (or see 
also \cite{hep-th/0103233}). The meaning of the mixing between 
Dirac brane currents and $m\exp(\t\beta B_2)$ will become clear below.

Once again, the modified CS term has precisely the correct coefficients to
ensure that the EOM can be written in terms of the gauge-invariant field
strengths. Of course, this fact is related to gauge invariance of the action.
If we take gauge transformations $\delta C_1=d\chi_0$, 
$\delta C_3=d\chi_2-\t\beta\chi_0 H_3$, 
\bea
\delta S &=& \frac{1}{2\kappa_0^2}\int\left[ \chi_0\left(d\star j_1+\t\beta
\star j_3\w H_3\right)\right.\label{varyIIA}\\
&&\left. +\chi_2\w\left(d\star j_3+\frac 12 \beta_0\t\beta
d\t F_4\w H_3-\frac 12 \t\beta d\star J_6\w H_3\right)\right] .\nonumber
\eea
Equation (\ref{nonconservation2}) implies that the $\chi_0$ terms cancel;
the $\chi_2$ terms also require the Bianchi identity and the divergence of
the Dirac brane current. The gauge invariance in fact ensures that the
integral over the potentials $C_1,C_3$ is well defined.

\subsection{The massive IIA theory from Dirac branes}

It is well known \cite{hep-th/9510017,hep-th/9510169} that D8-branes
are a source for the mass parameter of the Romans massive supergravity,
which is quantized in units of the D8-brane charge \cite{hep-th/9510227}.
As a review, since D8-branes are codimension one, $m$ is piecewise constant 
and jumps by one unit of D8-brane charge at the position of each D8-brane,
for example on a $S^1/Z_2$ orientifold. We conjecture that, in fact, the
currents of the associated Dirac 9-branes describe all the additional 
couplings of the massive IIA theory, whether described as a pure supergravity
or as the massless IIA theory in the presence of D8-branes and O8-planes.
Here, we present evidence in favor of this conjecture, point out some 
remaining questions, and illuminate consequences.

Start by considering the Bianchi identity $d\t F_0=\star j_9$ on the interval
transverse to a set of parallel D8-branes. If we define 
$\t F_0=m+\beta_{-2}\star J_{10}$, where $m$ is the bare
mass parameter, we see
that $m$ jumps by $\pm\mu_8$ at the location of each reference brane but
is otherwise constant. Meanwhile, $d\star J_{10}=\star(j_9-j_9^*)$, so 
$\star J_{10}$ is also a step function equal to $\pm\mu_8$ between the 
physical and reference branes. On the $S^1/Z_2$ orientifold, $m=0$ if half
of the reference D8-branes are coincident with each O8-plane. Alternately,
we can generate the massive supergravity by considering the massless IIA
theory on a circle. Then imagine an instantonic process in which a 
D8/$\bar{\textnormal{D}}$8-brane pair appears transverse to the circle, 
and then the
brane and antibrane move in opposite directions around the circle before
reannihilating.  This process leaves behind a closed Dirac 9-brane
extending around the entire circle and filling spacetime 
(there is a reference brane/antibrane pair
at the point of initial pair creation, which has no net 
effect).\footnote{\cite{hep-th/9510227} also suggests this brane nucleation 
process as a way to generate the bare mass parameter $m$.}
In both these cases, $\t F_0=-\beta_{0}N\mu_8$ for integer $N$, or 
$\star J_{10}=-\beta_{0}\t F_0$.

Since the Dirac 9-brane fills spacetime, it is natural to treat the WZ 
couplings on its currents as part of the bulk action. Ignoring any worldvolume
gauge fields, the Dirac brane currents are given by
\bea
\star J_8 &=& -\beta_{0}\eta\t F_0 B_2,\quad
\star J_6=-\frac{\beta_{0}\t F_0}{2} B_2^2 ,\nonumber\\ 
\star J_4&=&-\frac{\beta_{0}\eta\t F_0}{6} B_2^3 ,\quad
\star J_2=-\frac{\beta_{0}\t F_0}{24} B_2^4 , \nonumber\\
\star J_0&=&-\frac{\beta_{0}\eta\t F_0}{120} B_2^5 .\label{massiveJ}
\eea
As a result, the field strengths become (with $\eta=-\t\beta$ from the anomaly
inflow)
\beq{massiveF}
\t F_2 =dC_1 +\t\beta \t F_0 B_2 ,\quad \t F_4 = dC_3+\t\beta C_1\w H_3
+\frac 12 \t F_0 B_2^2 ,
\eeq
standard for the massive supergravity. In terms of these field strengths,
the action (\ref{actionIIA}) in the absence of D-branes becomes
\bea
S_{IIA}&=& -\frac{1}{2\kappa_0^2} \int \left[ \frac 12 \t F_0\w\star\t F_0
+\frac 12 \t F_2\w\star\t F_2+\frac 12 \t F_4\w\star\t F_4\right.\nonumber\\
&&\left. +\frac 12 \beta_0
\t\beta C_3\w dC_3\w H_3 +\frac{1}{2}\beta_0\t\beta\t F_0 C_3\w B_2^2\w H_3 
\right.\nonumber\\
&&\left.
+\frac{1}{40} \beta_0\t\beta \t F_0^2 B_2^5 \right] .\label{massiveIIA}
\eea
This precisely matches the action for Romans massive supergravity given
above (there, $\t F_0\to m$).
In particular, if we consider the 10D spacetime as the boundary of an
11D spacetime, the last three terms together are 
$(\beta_0\t\beta/2)\t F_4^2 H_3$ integrated in 11D, as expected. So we see
that the action for pure massive IIA supergravity follows from Dirac
brane currents. If we include both a bare mass parameter $m$ and the
Dirac branes (as is necessary generically in the presence of D8-branes),
we have $\star J_{10}=-\beta_0(\t F_0-m)$, which adjusts the coefficients 
in equation (\ref{massiveJ}). With this change, the field strength definitions
(\ref{massiveF}) still hold, and the mixed terms in the action 
(\ref{actionIIA}) are unchanged in terms of the physical $\t F_0$.

As a short aside, the last term in (\ref{massiveIIA}) 
is the sum of the $(\star J)^2$-type terms. Therefore, these
terms have physical consequences even in the classical theory. They are not
purely topological.

The interpretation of the mass parameter as the consequence of Dirac brane
currents raises a question about gauge transformations in the massive IIA 
theory. Since the field strengths (\ref{massiveF}) contain the potential $B_2$,
they are gauge invariant only if the RR potentials also transform under
the $B_2$ gauge transformations. This still seems to be case when $\t F_0$
arises from D8-instantons as above. However, in the case where D8-branes are
present, the Dirac brane currents depend on $\b\F$, the extension of
the gauge-invariant D8-brane field strength. In that case, the $B_2$ gauge
transformation is compensated by a corresponding transformation of $\b A_1$.
When $\t F_0$ is a mix of bare mass parameter $m$ and $\star J_{10}$ in the
presence of D8-branes, it seems
that the gauge transformation of $C$ would depend only on $m$, not the 
physical flux $\t F_0$.

Additional terms in the WZ action arise for D-branes in backgrounds with
$\t F_0\neq 0$, which \cite{hep-th/9603123,hep-th/9604119}  
demonstrated for type IIA D-branes via T duality. These take the form
\beq{extraWZ}
S_{WZ,\t F_0} = \frac{\mu_p}{(p/2+1)!} \int_\M [\t F_0]\, \omega_{p+1}, \eeq
where $\omega_{p+1}$ is the Chern-Simons form defined by 
$\h d\omega_{p+1}=(\h dA_1)^{(p+2)/2}$. We argue that these terms follow
naturally from the Dirac brane current $J_0$ and see that
these and other related WZ terms appear for all the IIA D-branes. 
This Dirac brane current for a D$(2n-2)$-brane is 
\bea
J_0 &=& J_{\N,\b\F^n/n!}=\sum_{\ell}
\frac{\mu_{2n-2}(2\pi\alpha')^\ell}{\ell!(n-\ell)!}\int_\N \b d\b\omega_{2\ell-1}
\w [B_2]^{n-\ell}\delta^{10}(x,Y)\nonumber\\ &=& -\sum_{\ell}
\frac{(2\pi\alpha')^\ell}{\ell!(n-\ell)!}\left[
j_{\M,\omega_{2\ell-1}[B_2]^{n-\ell}} -j_{\M^*,\omega^*_{2\ell-1}[B_2]^{n-\ell}}\right.\nonumber\\ 
&&\left.\vphantom{j_\M}+(n-\ell)
J_{\N,\b\omega_{2\ell-1}[H_3][B_2]^{n-\ell-1}}+\star d\star 
J_{\N,\b\omega_{2\ell-1}[B_2]^{n-\ell}}\right] .\label{J0current}
\eea
The first term, when substituted into the last term of the
action (\ref{actionIIA}),
is equivalent to a contribution 
$S_{WZ}\propto [\t F_0]\omega_{2\ell-1}[B_2]^{n-\ell}$
to the D-brane's WZ action, where the $\ell=n$ term reproduces
(\ref{extraWZ}). The remaining terms are a similar coupling on the
reference brane, a Dirac brane coupling involving $H_3$ flux, and (after
integration by parts) a contact term between the Dirac brane and any 
D8-brane.\footnote{While we presented (\ref{J0current}) in terms of a
Dirac brane, using the Chern kernel gives the same contributions with the
latter two terms extended over spacetime.}

This observation suggests that the additional WZ terms (\ref{extraWZ}) are
actually properly interpreted as Dirac brane couplings and should include
additional couplings to $B_2$. They reduce to (\ref{extraWZ}) in the
absence of $H_3$ flux and D8-branes; the additional term on the reference
brane does not contribute to the D-brane gauge field EOM because the
reference potential is fixed. Furthermore, (\ref{J0current}) immediately
implies that all type IIA D-branes have such couplings. So the action
(\ref{actionIIA}) leads to a prediction for D-branes in massive IIA
backgrounds. However, there is a puzzle. The $\t F_0\star J_0$ term in
the action is multiplied by a factor of $1/2$, so we have actually found
half of the WZ term suggested by T duality. A possible resolution is to note
that this term is analogous to the integration by parts of $C_{p+1}\star j_{p+1}$,
suggesting that we should include an extra $\t F_0\star J_0/2$ or perhaps
$m\star J_0/2$ (since the bare mass parameter depends on reference D8-branes)
already in action (\ref{democraticIIA}). The difficulty, of course, is that
adding this term to (\ref{actionIIA}) spoils agreement with the known action
for the massive IIA theory. Alternately, there could be a subtlety in the 
derivation of these WZ terms by T duality in 
\cite{hep-th/9603123,hep-th/9604119}. Specifically, it may be that any IIB
background dual to an allowed brane in the massive IIA theory involves an
orientifold, and the T duality rules in the presence of orientifolds can
introduce factors of 2 in bulk fields in comparison to T duality 
without orientifolds. This could change the weight of the new WZ terms.
Resolving this puzzle is a question for the future.

%%%%%%
\section{Discussion and future directions}\label{s:future}

We gave a brief introduction to the description of D-brane WZ actions
as they appear in the bulk supergravity action through D-brane and Dirac 
brane currents. We initially went through an anomaly inflow argument 
in terms of the nonconservation of D-brane currents and CS and transgression
terms in the EOM and Bianchi identities for the RR fields (for a generalized
version of the 10D type II supergravities). This discussion made explicit 
several points that are implicit in the literature, including the allowed
sign conventions for the type II supergravity actions --- assuming there is
a common sign choice for transgression terms in the field strengths, there
is one independent choice of sign on a magnetic current in the Bianchi 
identities. We then showed that reproducing the EOM in terms of gauge-invariant
field strengths requires a brane-induced modification to CS terms in the
action for the RR gauge fields. Inclusion of $\alpha'$ corrections in the 
brane currents, such as the A-roof genus terms, is explained implicitly in 
\cite{hep-th/0406083}, but an explicit description may be interesting.

Our main concern was to give actions for the IIB and IIA supergravities
with D-brane sources. As a preliminary, we explained how to integrate over
gauge-patched RR-sector potentials (or with similar gauge transformations).
A critical feature is the excision of the reference magnetic currents (around
which the potentials are patched), which leads to new surface terms.
We left any subtleties surrounding brane intersections to the future, 
and these may be important in reproducing additional anomaly terms found in
\cite{hep-th/0406083}.

Up to those higher $\alpha'$ corrections and brane intersection terms, 
we then reproduced the results of \cite{hep-th/0406083} for the action of
IIB supergravity with D-branes by dualizing the democratic formulation of
the theory. We uncovered the same brane-induced modification to the CS term 
as well as couplings between Dirac brane currents. We also showed that both
the standard CS term and the brane-induced term are necessary for invariance
of the action under gauge transformations of the RR potentials. With an eye
toward dimensional reduction and other applications where accounting for 
degrees of freedom is important, we further dualized the action, keeping only
half the degrees of freedom of the self-dual 4-form potential.

Finally, we presented the action of IIA supergravity with D-branes, including
the Romans mass parameter. This has a similarly modified CS term as the IIB
supergravity along with current-current couplings for Dirac brane currents,
though the latter are mixed with additional terms including the mass
parameter. We then demonstrated how Dirac brane currents carried by a 
Dirac 9-brane reproduce the entire action of the massive IIA theory without
D-branes. In fact, Dirac brane currents associated with other D-branes 
reproduce the form of additional WZ couplings on those branes in the massive
theory, which had been found by T duality \cite{hep-th/9603123,hep-th/9604119}.
However, these results raise some questions: In the Romans supergravity, 
RR potentials transform under the Neveu-Schwarz--Neveu-Schwarz (NSNS) 
gauge transformations, but should
the Dirac brane worldvolume potentials absorb those gauge transformations
instead? And what is the origin of the factor of 2 difference between
the new WZ couplings in massive supergravity as deduced from the Dirac brane
currents as opposed to T duality?

As we discussed in the introduction, Dirac's formalism separates the
brane and gauge field degrees of freedom. Identifying the correct degrees
of freedom is a critical task in a number of applications, including
determining the effective field theory of a dimensional reduction, for
example. (In fact, \cite{arXiv:1609.05904} used a variation on Dirac's
formalism to address the effective field theory of D3-branes in flux 
compactifications.) We will return to this issue in a forthcoming
companion paper \cite{freynew}, compiling useful formulae for the dimensional
reduction of branes and fluxes. The companion paper will also describe
several magnetic brane configurations, including examples of smoothly 
distributed magnetic monopole charge in electrodynamics and branes ending on
branes in string theory.

Looking further afield, other types of magnetic couplings and 
magnetically charged branes in string theory are targets for this analysis. 
First, a stack of D-branes carries a non-Abelian gauge theory, so extending
our results to non-Abelian worldvolume $F_2$ and to include noncommuting
worldvolume positions, which appear in the CS action of \cite{Myers:1999ps},
is an important task. Further, Lechner and co-workers
\cite{hep-th/0107061,hep-th/0203238,hep-th/0302108,hep-th/0402078}
and Bandos \textit{et al}. \cite{Bandos:1997gd}
have considered type IIA NS5-branes and M2- and M5-branes, and type IIB 
NS5-branes would be a logical next step. An important issue, as we noted,
is understanding the supergravity action in the presence of both D-branes
and NS5-branes, particularly the ``extra'' CS term sometimes included in
the IIB supergravity action and which is topological except in the presence 
of both D5- and NS5-branes. We also now know of numerous types of exotic
branes in string theory (along with KK monopoles), many of which also 
presumably have associated Dirac brane currents. How do these affect 
any effective gravitational and gauge action? We leave these intriguing
questions to the future.

\acknowledgments
ARF thanks K.~Dasgupta and R.~Danos for helpful discussions and K.~Lechner
and D.~Sorokin for correspondence.
ARF is supported by the Natural Sciences and
Engineering Research Council of Canada Discovery Grant program,
grants 2015-00046 and 2020-00054.  Part
of this work was carried out while visiting the Perimeter Institute
for Theoretical Physics. Research at Perimeter Institute is supported
by the Government of Canada through the Department of Innovation,
Science and Economic Development and by the Province of Ontario
through the Ministry of Research and Innovation.

%%%%%%
\appendix
\section{Conventions}\label{a:conventions}
We follow the conventions of \cite{Polchinski:1998rr} for forms, 
in particular taking $\epsilon_{0\cdots D} =+\sqrt{|g|}$ and 
\beq{star}
(\star A_p)_{\mu_1\cdots\mu_{D-p}} = \frac{1}{p!}\epsilon_{\mu_1\cdots\mu_{D-p}}
{}^{\nu_1\cdots\nu_p} A_{\nu_1\cdots\nu_p}.\eeq
This leads to $\star\star A_p=(-1)^{p(D-p)+s}A_p$ in a space of signature $s$
($=0$ Euclidean, $=1$ Lorentzian).  Other useful identities include
\bea
\star\left(A_p\w\star B_q\right)_{\mu_1\cdots\mu_{q-p}} &=&\frac{(-1)^{q(D-q)+s}
}{p!}
B_{\mu_1\cdots\mu_{q-p}\nu_1\cdots\nu_p}A^{\nu_1\cdots\nu_p}\nonumber\\
&\equiv&(-1)^{q(D-q)+s}(B_q\cdot A_p)_{\mu_1\cdots\mu_{q-p}},\nonumber\\
(\star d\star A_p)_{\mu_1\cdots\mu_{p-1}}&=&(-1)^{p(D-p+1)+s+1}\Del^\nu 
A_{\nu\mu_1\cdots\mu_{p-1}}.\label{idents}
\eea
At times we will consider formal sums of forms of different ranks and 
products of such sums; integrating over a manifold of some particular
dimension picks out only the form of that rank.  We will suppress 
explicit wedge signs for formulae in text or sub/superscripts for notational
convenience.

We typically use capital letters to denote the embedding coordinates of branes,
taking $X$ for physical branes and $Y$ for Dirac branes. Exterior derivatives
without accents are spacetime derivatives, hatted exterior derivatives are
along physical branes, and barred exterior derivatives are along Dirac branes.
We denote the pullback of a spacetime form to a brane worldvolume with
square brackets, so
\beq{pullback} [A_p] = \frac{1}{p!} A_{\mu_1\cdots\mu_p}\h d X^{\mu_1}\w\cdots
\h dX^{\mu_p}\eeq
(and similarly for Dirac branes).  It is worth noting that
$[dA]=\h d[A]$ because partial derivatives of the form 
$\del^n X^\mu/\del\xi^{a_1}\cdots\del\xi^{a_n}$ commute.

When integrating forms, we must specify an orientation for the integration
measure in order to determine all signs.  In particular, we choose a 
worldvolume coordinate transverse to the worldvolume boundary as the last
coordinate in the integration measure.  Thus, the integral over a $(p+1)$-brane
$\M$ is
\beq{measure}
\int_\M \h dA_p =(-1)^p \int_{\del\M} A_p .\eeq

\section{Gauge invariance of the extended theory without branes}
\label{a:invariance}

In this appendix, we will determine an action principle for the 
general gauge theory of forms discussed in sections \ref{s:currents}
and \ref{s:actions} including relationships between the constants that appear 
in the EOM (\ref{eomF2}) and Bianchi identities (\ref{bianchiF2}).  
We will be forced to consider the gauge invariance of the potentials $C_{p+1}$; 
this is of course related to our discussion of monopole branes and 
the Dirac brane formalism, but we consider here the situation with no brane
sources and globally defined gauge transformations (not gauge patching).

We start with the action
\bea
S&=&\int \sum_p \left[(-1)^{p(D-p)+1}\frac 12 \t F_{p+2}\w\star\t F_{p+2}\right.
\nonumber\\
&&\left. +\sum_r\gamma_{p,r} C_{p+1}\w \t F_{D-p-r-2}\w H_{r+1}\right]
\label{action2}\eea
where the sums run over the values of $p,r$ corresponding to extant 
field strengths and potentials. The latter sum is the Chern-Simons action.
The field strength is defined
\beq{transgression2}
\t F_{p+2}=dC_{p+1}+\sum_r \t\beta_{p,r}C_{p-r+1}\w H_{r+1} ;
\eeq
$H_{r+1}=dB_r$ are another set of exact field strengths whose kinetic terms
we ignore here. $\t\beta_{p,r}$ and
$\gamma_{p,r}$ are constants, which we take to vanish for values of $p,r$ 
where the corresponding potentials and field strengths do not exist.

Gauge invariance places constraints on these constants.  Consider first
gauge invariance of the field strength $\t F_{p+2}$ with gauge transformations
\beq{gauge1}
\delta C_{p+1} = d\Lambda_p-\sum_r \t\beta_{p,r}\Lambda_{p-r}\w H_{r+1} .\eeq
Then 
\beq{gauge2}
\delta\t F_{p+2} = -\sum_{r,l} \t\beta_{p,r}\t\beta_{p-r,l}\Lambda_{p-r-l}\w
H_{l+1}\w H_{r+1}\eeq
with some cancellation occurring as is familiar in 10-dimensional supergravity;
the remaining terms vanish in that case because there is only one additional
field strength $H_3$. In general, these terms cannot be cancelled by extending
the gauge transformations (\ref{gauge1}). Instead, these terms must cancel
among themselves.  In the sum, each form combination 
$\Lambda_{p-r-l}H_{l+1}H_{r+1}$ appears twice, leading to the constraint
\beq{gaugeconstraint1}
\t\beta_{p,r}\t\beta_{p-r,l}+(-1)^{(r+1)(l+1)}\t\beta_{p,l}\t\beta_{p-l,r}=0 .
\eeq
We can also see this constraint in the requirement that the Bianchi identity
be written in terms of gauge-invariant variables.  Differentiating
(\ref{transgression2}), we have
\bea
d\t F_{p+2} &=& \sum_r\t\beta_{p,r} dC_{p-r+1} \w H_{r+1}\\
&=& \sum_r\t\beta_{p,r}\t F_{p-r+2}\w H_{r+1}\nonumber\\
&&-\sum_{r,l}\t\beta_{p,r}\t\beta_{p-r,l}C_{p-r-l+1}\w H_{l+1}\w H_{r+1} .\nonumber\eea
The additional undesired terms are precisely those given in (\ref{gauge2}) 
with the substitution $\Lambda_{p-r-l}\to C_{p-r-l+1}$, so they also vanish
when (\ref{gaugeconstraint1}) is satisfied. Additionally, the integrability
condition coming from the exterior derivative of the Bianchi identity is
\bea
0&=&\sum_r \t\beta_{p,r}d\t F_{p-r+2}\w H_{r+1}\label{bianchiintegrability2}\\
&=& \sum_{r,l}\t\beta_{p,r}\t\beta_{p-r,l}\t F_{p-r-l+2}\w H_{l+1}\w H_{r+1},\nonumber
\eea
which is again satisfied whenever (\ref{gaugeconstraint1}) is satisfied.

The gauge variation of the action (\ref{action2}) under (\ref{gauge1}) is
\bea
\delta S &=& -\int \sum_{p,r,l} \gamma_{p,r}\left[(-1)^p \t\beta_{D-p-r-4,l}
\Lambda_p\w \t F_{D-p-r-l-2}\right.\nonumber\\
&&\left. +(-1)^{(l+1)(D-p-r)}\t\beta_{p,l}\Lambda_{p-l}\w
\t F_{D-p-r-2}\right]\nonumber\\
&&\w H_{l+1}\w H_{r+1} .\label{gauge3}\eea
If we shift the sum over $p$ in the second term, the gauge variation vanishes
if 
\bea
0&=&(-1)^p\gamma_{p,r}\t\beta_{D-p-r-4,l}+(-1)^{p+(r+1)(l+1)}\gamma_{p,l}
\t\beta_{D-p-l-4,r}\nonumber\\
&&+(-1)^{(l+1)(D-p-r-l)}\gamma_{p+l,r}\t\beta_{p+l,l}\nonumber\\
&&+(-1)^{(r+1)(D-p-r-1)}\gamma_{p+r,l}\t\beta_{p+r,r}\label{gaugeconstraint2}\eea
including the fact that the double sum over $r,l$ duplicates terms.
Meanwhile, the EOM is
\bea
d\star\t F_{p+2} &=& -\sum_r (-1)^{D+r(p+1)}\t\beta_{p+r,r}(\star\t F_{p+r+2})\w
H_{r+1}\nonumber\\
&& +\sum_r \left((-1)^{pD}\gamma_{p,r}+(-1)^{p(p+r)}\gamma_{D-p-r-4,r}\right)
\nonumber\\
&&\times \t F_{D-p-r-2}\w H_{r+1}\label{EOMaction}
\eea
if
\bea
0&=&\sum_{r,l} \left[(-1)^{D-r-l-pl-1}\gamma_{D-p-r-l-4,r}\t\beta_{p+l,l}
\right.\label{gaugeconstraint3}\\
&&\left.-\gamma_{D-p-r-4,r}\t\beta_{D-p-r-4,l}\right]C_{D-p-r-l-3}\w H_{l+1}\w H_{r+1}
\nonumber \eea
(so the EOM is in terms of the gauge-invariant field strength).
As usual, we can rewrite (\ref{gaugeconstraint3}) to account for duplication
in the sum and rename $p\leftrightarrow D-p-r-l-4$ for comparison to 
(\ref{gaugeconstraint2}).  For consistency, we find
\beq{gaugeconstraint4}
(-1)^p\gamma_{p,r}\t\beta_{D-p-r-4,l} +(-1)^{(l+1)(D-p-r-l)}\gamma_{p+l,r}\t\beta_{p+l,l}
=0 .\eeq

We also see that the coefficients in the EOM as defined in (\ref{eomF2})
relate to the coefficients in the action as follows:
\bea
\alpha_{p,r}&=&(-1)^{D+r(p+1)+1}\t\beta_{p+r,r} ,\nonumber\\
\t\alpha_{p,r}&=&(-1)^{pD}\gamma_{p,r}+(-1)^{p(p+r)}\gamma_{D-p-r-4,r} .
\label{coeffs}\eea
(If $\t F_{p+r+2=D/2}$ is self-dual, its kinetic term is halved, so $\alpha_{p,r}$
takes half the value given above. Meanwhile, $\t\alpha_{p,r}$ for self-dual
$\t F_{p+2=D/2}$ is doubled for the same reason.)
The integrability condition from the exterior derivative of the EOM 
(\ref{eomF2}) is
\bea
0&=&\sum_{r,l}\left\{\alpha_{p,r}\alpha_{p+r,l}\star\t F_{p+r+l+2}+\left(\alpha_{p,r}
\t\alpha_{p+r,l}\vphantom{\t\beta}\right.\right.\label{eomintegrability2}\\
&&\left.\left.+\t\alpha_{p,r}\t\beta_{D-p-r-4,l}\right)\t F_{D-p-r-l-2}\right\}
\w H_{l+1}\w H_{r+1} .\nonumber\eea
It is straightforward (but somewhat tedious) to show that this is satisfied
as long as (\ref{gaugeconstraint1}) and (\ref{gaugeconstraint4}) 
are satisfied for coefficients (\ref{coeffs}).

\bibliography{diracbranes1}

\begin{thebibliography}{50}%
\makeatletter
\providecommand \@ifxundefined [1]{%
 \@ifx{#1\undefined}
}%
\providecommand \@ifnum [1]{%
 \ifnum #1\expandafter \@firstoftwo
 \else \expandafter \@secondoftwo
 \fi
}%
\providecommand \@ifx [1]{%
 \ifx #1\expandafter \@firstoftwo
 \else \expandafter \@secondoftwo
 \fi
}%
\providecommand \natexlab [1]{#1}%
\providecommand \enquote  [1]{``#1''}%
\providecommand \bibnamefont  [1]{#1}%
\providecommand \bibfnamefont [1]{#1}%
\providecommand \citenamefont [1]{#1}%
\providecommand \href@noop [0]{\@secondoftwo}%
\providecommand \href [0]{\begingroup \@sanitize@url \@href}%
\providecommand \@href[1]{\@@startlink{#1}\@@href}%
\providecommand \@@href[1]{\endgroup#1\@@endlink}%
\providecommand \@sanitize@url [0]{\catcode `\\12\catcode `\$12\catcode
  `\&12\catcode `\#12\catcode `\^12\catcode `\_12\catcode `\%12\relax}%
\providecommand \@@startlink[1]{}%
\providecommand \@@endlink[0]{}%
\providecommand \url  [0]{\begingroup\@sanitize@url \@url }%
\providecommand \@url [1]{\endgroup\@href {#1}{\urlprefix }}%
\providecommand \urlprefix  [0]{URL }%
\providecommand \Eprint [0]{\href }%
\providecommand \doibase [0]{http://dx.doi.org/}%
\providecommand \selectlanguage [0]{\@gobble}%
\providecommand \bibinfo  [0]{\@secondoftwo}%
\providecommand \bibfield  [0]{\@secondoftwo}%
\providecommand \translation [1]{[#1]}%
\providecommand \BibitemOpen [0]{}%
\providecommand \bibitemStop [0]{}%
\providecommand \bibitemNoStop [0]{.\EOS\space}%
\providecommand \EOS [0]{\spacefactor3000\relax}%
\providecommand \BibitemShut  [1]{\csname bibitem#1\endcsname}%
\let\auto@bib@innerbib\@empty
%</preamble>
\bibitem [{\citenamefont {Wu}\ and\ \citenamefont
  {Yang}(1976{\natexlab{a}})}]{Wu:1976ge}%
  \BibitemOpen
  \bibfield  {author} {\bibinfo {author} {\bibfnamefont {Tai~Tsun}\
  \bibnamefont {Wu}}\ and\ \bibinfo {author} {\bibfnamefont {Chen~Ning}\
  \bibnamefont {Yang}},\ }\bibfield  {title} {\enquote {\bibinfo {title}
  {{Dirac Monopole Without Strings: Monopole Harmonics}},}\ }\href {\doibase
  10.1016/0550-3213(76)90143-7} {\bibfield  {journal} {\bibinfo  {journal}
  {Nucl. Phys.}\ }\textbf {\bibinfo {volume} {B107}},\ \bibinfo {pages} {365}
  (\bibinfo {year} {1976}{\natexlab{a}})}\BibitemShut {NoStop}%
%%CITATION = NUPHA,B107,365;%%
\bibitem [{\citenamefont {Wu}\ and\ \citenamefont
  {Yang}(1976{\natexlab{b}})}]{Wu:1976qk}%
  \BibitemOpen
  \bibfield  {author} {\bibinfo {author} {\bibfnamefont {Tai~Tsun}\
  \bibnamefont {Wu}}\ and\ \bibinfo {author} {\bibfnamefont {Chen~Ning}\
  \bibnamefont {Yang}},\ }\bibfield  {title} {\enquote {\bibinfo {title}
  {{Dirac's Monopole Without Strings: Classical Lagrangian Theory}},}\ }\href
  {\doibase 10.1103/PhysRevD.14.437} {\bibfield  {journal} {\bibinfo  {journal}
  {Phys. Rev.}\ }\textbf {\bibinfo {volume} {D14}},\ \bibinfo {pages}
  {437--445} (\bibinfo {year} {1976}{\natexlab{b}})},\ \bibinfo {note}
  {[,509(1976)]}\BibitemShut {NoStop}%
%%CITATION = PHRVA,D14,437;%%
\bibitem [{\citenamefont {Bergshoeff}\ \emph {et~al.}(2001)\citenamefont
  {Bergshoeff}, \citenamefont {Kallosh}, \citenamefont {Ortin}, \citenamefont
  {Roest},\ and\ \citenamefont {Van~Proeyen}}]{hep-th/0103233}%
  \BibitemOpen
  \bibfield  {author} {\bibinfo {author} {\bibfnamefont {Eric}\ \bibnamefont
  {Bergshoeff}}, \bibinfo {author} {\bibfnamefont {Renata}\ \bibnamefont
  {Kallosh}}, \bibinfo {author} {\bibfnamefont {Tomas}\ \bibnamefont {Ortin}},
  \bibinfo {author} {\bibfnamefont {Diederik}\ \bibnamefont {Roest}}, \ and\
  \bibinfo {author} {\bibfnamefont {Antoine}\ \bibnamefont {Van~Proeyen}},\
  }\bibfield  {title} {\enquote {\bibinfo {title} {{New formulations of D = 10
  supersymmetry and D8 - O8 domain walls}},}\ }\href {\doibase
  10.1088/0264-9381/18/17/303} {\bibfield  {journal} {\bibinfo  {journal}
  {Class. Quant. Grav.}\ }\textbf {\bibinfo {volume} {18}},\ \bibinfo {pages}
  {3359--3382} (\bibinfo {year} {2001})},\ \Eprint
  {http://arxiv.org/abs/hep-th/0103233} {arXiv:hep-th/0103233 [hep-th]}
  \BibitemShut {NoStop}%
%%CITATION = HEP-TH/0103233;%%
\bibitem [{\citenamefont {Bandos}\ \emph {et~al.}(2004)\citenamefont {Bandos},
  \citenamefont {Nurmagambetov},\ and\ \citenamefont
  {Sorokin}}]{Bandos:2003et}%
  \BibitemOpen
  \bibfield  {author} {\bibinfo {author} {\bibfnamefont {Igor~A.}\ \bibnamefont
  {Bandos}}, \bibinfo {author} {\bibfnamefont {Alexei~J.}\ \bibnamefont
  {Nurmagambetov}}, \ and\ \bibinfo {author} {\bibfnamefont {Dmitri~P.}\
  \bibnamefont {Sorokin}},\ }\bibfield  {title} {\enquote {\bibinfo {title}
  {{Various faces of type IIA supergravity}},}\ }\href {\doibase
  10.1016/j.nuclphysb.2003.10.036} {\bibfield  {journal} {\bibinfo  {journal}
  {Nucl. Phys.}\ }\textbf {\bibinfo {volume} {B676}},\ \bibinfo {pages}
  {189--228} (\bibinfo {year} {2004})},\ \Eprint
  {http://arxiv.org/abs/hep-th/0307153} {arXiv:hep-th/0307153 [hep-th]}
  \BibitemShut {NoStop}%
%%CITATION = HEP-TH/0307153;%%
\bibitem [{\citenamefont {Dall'Agata}\ \emph {et~al.}(1998)\citenamefont
  {Dall'Agata}, \citenamefont {Lechner},\ and\ \citenamefont
  {Tonin}}]{DallAgata:1998ahf}%
  \BibitemOpen
  \bibfield  {author} {\bibinfo {author} {\bibfnamefont {Gianguido}\
  \bibnamefont {Dall'Agata}}, \bibinfo {author} {\bibfnamefont {Kurt}\
  \bibnamefont {Lechner}}, \ and\ \bibinfo {author} {\bibfnamefont {Mario}\
  \bibnamefont {Tonin}},\ }\bibfield  {title} {\enquote {\bibinfo {title} {{D =
  10, N = IIB supergravity: Lorentz invariant actions and duality}},}\ }\href
  {\doibase 10.1088/1126-6708/1998/07/017} {\bibfield  {journal} {\bibinfo
  {journal} {JHEP}\ }\textbf {\bibinfo {volume} {07}},\ \bibinfo {pages} {017}
  (\bibinfo {year} {1998})},\ \Eprint {http://arxiv.org/abs/hep-th/9806140}
  {arXiv:hep-th/9806140 [hep-th]} \BibitemShut {NoStop}%
%%CITATION = HEP-TH/9806140;%%
\bibitem [{\citenamefont {Dirac}(1948)}]{Dirac:1948um}%
  \BibitemOpen
  \bibfield  {author} {\bibinfo {author} {\bibfnamefont {Paul A.~M.}\
  \bibnamefont {Dirac}},\ }\bibfield  {title} {\enquote {\bibinfo {title} {{The
  Theory of magnetic poles}},}\ }\href {\doibase 10.1103/PhysRev.74.817}
  {\bibfield  {journal} {\bibinfo  {journal} {Phys. Rev.}\ }\textbf {\bibinfo
  {volume} {74}},\ \bibinfo {pages} {817--830} (\bibinfo {year}
  {1948})}\BibitemShut {NoStop}%
%%CITATION = PHRVA,74,817;%%
\bibitem [{\citenamefont {Brandt}\ and\ \citenamefont
  {Primack}(1977{\natexlab{a}})}]{Brandt:1977ks}%
  \BibitemOpen
  \bibfield  {author} {\bibinfo {author} {\bibfnamefont {Richard~A.}\
  \bibnamefont {Brandt}}\ and\ \bibinfo {author} {\bibfnamefont {Joel~R.}\
  \bibnamefont {Primack}},\ }\bibfield  {title} {\enquote {\bibinfo {title}
  {{Dirac monopole theory with and without strings}},}\ }\href {\doibase
  10.1103/PhysRevD.15.1175} {\bibfield  {journal} {\bibinfo  {journal} {Phys.
  Rev.}\ }\textbf {\bibinfo {volume} {D15}},\ \bibinfo {pages} {1175} (\bibinfo
  {year} {1977}{\natexlab{a}})}\BibitemShut {NoStop}%
%%CITATION = PHRVA,D15,1175;%%
\bibitem [{\citenamefont {Brandt}\ and\ \citenamefont
  {Primack}(1977{\natexlab{b}})}]{Brandt:1976hk}%
  \BibitemOpen
  \bibfield  {author} {\bibinfo {author} {\bibfnamefont {Richard~A.}\
  \bibnamefont {Brandt}}\ and\ \bibinfo {author} {\bibfnamefont {Joel~R.}\
  \bibnamefont {Primack}},\ }\bibfield  {title} {\enquote {\bibinfo {title}
  {{Avoiding Dirac's Veto in Monopole Theory}},}\ }\href {\doibase
  10.1103/PhysRevD.15.1798} {\bibfield  {journal} {\bibinfo  {journal} {Phys.
  Rev.}\ }\textbf {\bibinfo {volume} {D15}},\ \bibinfo {pages} {1798--1802}
  (\bibinfo {year} {1977}{\natexlab{b}})}\BibitemShut {NoStop}%
%%CITATION = PHRVA,D15,1798;%%
\bibitem [{\citenamefont {Teitelboim}(1986)}]{Teitelboim:1985yc}%
  \BibitemOpen
  \bibfield  {author} {\bibinfo {author} {\bibfnamefont {Claudio}\ \bibnamefont
  {Teitelboim}},\ }\bibfield  {title} {\enquote {\bibinfo {title} {{Monopoles
  of Higher Rank}},}\ }\href {\doibase 10.1016/0370-2693(86)90547-2} {\bibfield
   {journal} {\bibinfo  {journal} {Phys. Lett.}\ }\textbf {\bibinfo {volume}
  {167B}},\ \bibinfo {pages} {69--72} (\bibinfo {year} {1986})}\BibitemShut
  {NoStop}%
%%CITATION = PHLTA,167B,69;%%
\bibitem [{\citenamefont {Bandos}\ \emph {et~al.}(1998)\citenamefont {Bandos},
  \citenamefont {Berkovits},\ and\ \citenamefont {Sorokin}}]{Bandos:1997gd}%
  \BibitemOpen
  \bibfield  {author} {\bibinfo {author} {\bibfnamefont {Igor~A.}\ \bibnamefont
  {Bandos}}, \bibinfo {author} {\bibfnamefont {Nathan}\ \bibnamefont
  {Berkovits}}, \ and\ \bibinfo {author} {\bibfnamefont {Dmitri~P.}\
  \bibnamefont {Sorokin}},\ }\bibfield  {title} {\enquote {\bibinfo {title}
  {{Duality symmetric eleven-dimensional supergravity and its coupling to
  M-branes}},}\ }\href {\doibase 10.1016/S0550-3213(98)00102-3} {\bibfield
  {journal} {\bibinfo  {journal} {Nucl. Phys.}\ }\textbf {\bibinfo {volume}
  {B522}},\ \bibinfo {pages} {214--233} (\bibinfo {year} {1998})},\ \Eprint
  {http://arxiv.org/abs/hep-th/9711055} {arXiv:hep-th/9711055 [hep-th]}
  \BibitemShut {NoStop}%
%%CITATION = HEP-TH/9711055;%%
\bibitem [{\citenamefont {Lechner}\ and\ \citenamefont
  {Marchetti}(2001{\natexlab{a}})}]{hep-th/0103161}%
  \BibitemOpen
  \bibfield  {author} {\bibinfo {author} {\bibfnamefont {Kurt}\ \bibnamefont
  {Lechner}}\ and\ \bibinfo {author} {\bibfnamefont {Pieralberto}\ \bibnamefont
  {Marchetti}},\ }\bibfield  {title} {\enquote {\bibinfo {title} {{Dirac
  branes, characteristic currents and anomaly cancellations in five-branes}},}\
  }\bibfield  {booktitle} {\emph {\bibinfo {booktitle} {{Supersymmetry and
  quantum field theory. Proceedings, International Conference, SSQFT 2000,
  Kharkov, Ukraine, July 25-29, 2000}}},\ }\href {\doibase
  10.1016/S0920-5632(01)01542-0} {\bibfield  {journal} {\bibinfo  {journal}
  {Nucl. Phys. Proc. Suppl.}\ }\textbf {\bibinfo {volume} {102}},\ \bibinfo
  {pages} {94--99} (\bibinfo {year} {2001}{\natexlab{a}})},\ \bibinfo {note}
  {[,94(2000)]},\ \Eprint {http://arxiv.org/abs/hep-th/0103161}
  {arXiv:hep-th/0103161 [hep-th]} \BibitemShut {NoStop}%
%%CITATION = HEP-TH/0103161;%%
\bibitem [{\citenamefont {Lechner}\ and\ \citenamefont
  {Marchetti}(2001{\natexlab{b}})}]{hep-th/0007076}%
  \BibitemOpen
  \bibfield  {author} {\bibinfo {author} {\bibfnamefont {K.}~\bibnamefont
  {Lechner}}\ and\ \bibinfo {author} {\bibfnamefont {P.~A.}\ \bibnamefont
  {Marchetti}},\ }\bibfield  {title} {\enquote {\bibinfo {title} {{Interacting
  branes, dual branes, and dyonic branes: A Unifying Lagrangian approach in D
  dimensions}},}\ }\href {\doibase 10.1088/1126-6708/2001/01/003} {\bibfield
  {journal} {\bibinfo  {journal} {JHEP}\ }\textbf {\bibinfo {volume} {01}},\
  \bibinfo {pages} {003} (\bibinfo {year} {2001}{\natexlab{b}})},\ \Eprint
  {http://arxiv.org/abs/hep-th/0007076} {arXiv:hep-th/0007076 [hep-th]}
  \BibitemShut {NoStop}%
%%CITATION = HEP-TH/0007076;%%
\bibitem [{\citenamefont {Lechner}\ \emph {et~al.}(2002)\citenamefont
  {Lechner}, \citenamefont {Marchetti},\ and\ \citenamefont
  {Tonin}}]{hep-th/0107061}%
  \BibitemOpen
  \bibfield  {author} {\bibinfo {author} {\bibfnamefont {K.}~\bibnamefont
  {Lechner}}, \bibinfo {author} {\bibfnamefont {P.~A.}\ \bibnamefont
  {Marchetti}}, \ and\ \bibinfo {author} {\bibfnamefont {M.}~\bibnamefont
  {Tonin}},\ }\bibfield  {title} {\enquote {\bibinfo {title} {{Anomaly free
  effective action for the elementary M5 brane}},}\ }\href {\doibase
  10.1016/S0370-2693(01)01390-9} {\bibfield  {journal} {\bibinfo  {journal}
  {Phys. Lett.}\ }\textbf {\bibinfo {volume} {B524}},\ \bibinfo {pages}
  {199--207} (\bibinfo {year} {2002})},\ \Eprint
  {http://arxiv.org/abs/hep-th/0107061} {arXiv:hep-th/0107061 [hep-th]}
  \BibitemShut {NoStop}%
%%CITATION = HEP-TH/0107061;%%
\bibitem [{\citenamefont {Cariglia}\ and\ \citenamefont
  {Lechner}(2002)}]{hep-th/0203238}%
  \BibitemOpen
  \bibfield  {author} {\bibinfo {author} {\bibfnamefont {Marco}\ \bibnamefont
  {Cariglia}}\ and\ \bibinfo {author} {\bibfnamefont {Kurt}\ \bibnamefont
  {Lechner}},\ }\bibfield  {title} {\enquote {\bibinfo {title} {{NS5-branes in
  IIA supergravity and gravitational anomalies}},}\ }\href {\doibase
  10.1103/PhysRevD.66.045003} {\bibfield  {journal} {\bibinfo  {journal} {Phys.
  Rev.}\ }\textbf {\bibinfo {volume} {D66}},\ \bibinfo {pages} {045003}
  (\bibinfo {year} {2002})},\ \Eprint {http://arxiv.org/abs/hep-th/0203238}
  {arXiv:hep-th/0203238 [hep-th]} \BibitemShut {NoStop}%
%%CITATION = HEP-TH/0203238;%%
\bibitem [{\citenamefont {Lechner}\ and\ \citenamefont
  {Marchetti}(2003)}]{hep-th/0302108}%
  \BibitemOpen
  \bibfield  {author} {\bibinfo {author} {\bibfnamefont {Kurt}\ \bibnamefont
  {Lechner}}\ and\ \bibinfo {author} {\bibfnamefont {PierAlberto}\ \bibnamefont
  {Marchetti}},\ }\bibfield  {title} {\enquote {\bibinfo {title} {{Chern
  kernels and anomaly cancellation in M theory}},}\ }\href {\doibase
  10.1016/j.nuclphysb.2003.09.002} {\bibfield  {journal} {\bibinfo  {journal}
  {Nucl. Phys.}\ }\textbf {\bibinfo {volume} {B672}},\ \bibinfo {pages}
  {264--302} (\bibinfo {year} {2003})},\ \Eprint
  {http://arxiv.org/abs/hep-th/0302108} {arXiv:hep-th/0302108 [hep-th]}
  \BibitemShut {NoStop}%
%%CITATION = HEP-TH/0302108;%%
\bibitem [{\citenamefont {Lechner}(2004)}]{hep-th/0402078}%
  \BibitemOpen
  \bibfield  {author} {\bibinfo {author} {\bibfnamefont {Kurt}\ \bibnamefont
  {Lechner}},\ }\bibfield  {title} {\enquote {\bibinfo {title} {{Intersecting
  M2-branes and M5-branes}},}\ }\href {\doibase 10.1016/j.physletb.2004.03.056}
  {\bibfield  {journal} {\bibinfo  {journal} {Phys. Lett.}\ }\textbf {\bibinfo
  {volume} {B589}},\ \bibinfo {pages} {147--154} (\bibinfo {year} {2004})},\
  \Eprint {http://arxiv.org/abs/hep-th/0402078} {arXiv:hep-th/0402078 [hep-th]}
  \BibitemShut {NoStop}%
%%CITATION = HEP-TH/0402078;%%
\bibitem [{\citenamefont {Cariglia}\ and\ \citenamefont
  {Lechner}(2004)}]{hep-th/0406083}%
  \BibitemOpen
  \bibfield  {author} {\bibinfo {author} {\bibfnamefont {Marco}\ \bibnamefont
  {Cariglia}}\ and\ \bibinfo {author} {\bibfnamefont {Kurt}\ \bibnamefont
  {Lechner}},\ }\bibfield  {title} {\enquote {\bibinfo {title} {{Intersecting
  D-branes, Chern-kernels and the inflow mechanism}},}\ }\href {\doibase
  10.1016/j.nuclphysb.2004.08.023} {\bibfield  {journal} {\bibinfo  {journal}
  {Nucl. Phys.}\ }\textbf {\bibinfo {volume} {B700}},\ \bibinfo {pages}
  {157--182} (\bibinfo {year} {2004})},\ \Eprint
  {http://arxiv.org/abs/hep-th/0406083} {arXiv:hep-th/0406083 [hep-th]}
  \BibitemShut {NoStop}%
%%CITATION = HEP-TH/0406083;%%
\bibitem [{\citenamefont {Cownden}\ and\ \citenamefont
  {Frey}(2018)}]{arXiv:1807.07401}%
  \BibitemOpen
  \bibfield  {author} {\bibinfo {author} {\bibfnamefont {Brad}\ \bibnamefont
  {Cownden}}\ and\ \bibinfo {author} {\bibfnamefont {Andrew~R.}\ \bibnamefont
  {Frey}},\ }\bibfield  {title} {\enquote {\bibinfo {title} {{Variations on the
  Dirac string}},}\ }\href {\doibase 10.1103/PhysRevD.98.105013} {\bibfield
  {journal} {\bibinfo  {journal} {Phys. Rev.}\ }\textbf {\bibinfo {volume}
  {D98}},\ \bibinfo {pages} {105013} (\bibinfo {year} {2018})},\ \Eprint
  {http://arxiv.org/abs/1807.07401} {arXiv:1807.07401 [hep-th]} \BibitemShut
  {NoStop}%
%%CITATION = ARXIV:1807.07401;%%
\bibitem [{\citenamefont {Chern}(1944)}]{Chern1944}%
  \BibitemOpen
  \bibfield  {author} {\bibinfo {author} {\bibfnamefont {Shiing-Shen}\
  \bibnamefont {Chern}},\ }\bibfield  {title} {\enquote {\bibinfo {title} {{A
  Simple Intrinsic Proof of the Gauss-Bonnet Formula for Closed Riemannian
  Manifolds}},}\ }\href {\doibase 10.2307/1969302} {\bibfield  {journal}
  {\bibinfo  {journal} {Annals of Mathematics}\ }\textbf {\bibinfo {volume}
  {45}},\ \bibinfo {pages} {747--752} (\bibinfo {year} {1944})}\BibitemShut
  {NoStop}%
\bibitem [{\citenamefont {Chern}(1945)}]{Chern1945}%
  \BibitemOpen
  \bibfield  {author} {\bibinfo {author} {\bibfnamefont {Shiing-Shen}\
  \bibnamefont {Chern}},\ }\bibfield  {title} {\enquote {\bibinfo {title} {{On
  the Curvatura Integra in a Riemannian Manifold}},}\ }\href {\doibase
  10.2307/1969203} {\bibfield  {journal} {\bibinfo  {journal} {Annals of
  Mathematics}\ }\textbf {\bibinfo {volume} {46}},\ \bibinfo {pages} {674--684}
  (\bibinfo {year} {1945})}\BibitemShut {NoStop}%
\bibitem [{\citenamefont {Deser}\ \emph {et~al.}(1998)\citenamefont {Deser},
  \citenamefont {Gomberoff}, \citenamefont {Henneaux},\ and\ \citenamefont
  {Teitelboim}}]{Deser:1997se}%
  \BibitemOpen
  \bibfield  {author} {\bibinfo {author} {\bibfnamefont {Stanley}\ \bibnamefont
  {Deser}}, \bibinfo {author} {\bibfnamefont {A.}~\bibnamefont {Gomberoff}},
  \bibinfo {author} {\bibfnamefont {M.}~\bibnamefont {Henneaux}}, \ and\
  \bibinfo {author} {\bibfnamefont {C.}~\bibnamefont {Teitelboim}},\ }\bibfield
   {title} {\enquote {\bibinfo {title} {{P-brane dyons and electric magnetic
  duality}},}\ }\href {\doibase 10.1016/S0550-3213(98)00179-5} {\bibfield
  {journal} {\bibinfo  {journal} {Nucl. Phys.}\ }\textbf {\bibinfo {volume}
  {B520}},\ \bibinfo {pages} {179--204} (\bibinfo {year} {1998})},\ \Eprint
  {http://arxiv.org/abs/hep-th/9712189} {arXiv:hep-th/9712189 [hep-th]}
  \BibitemShut {NoStop}%
%%CITATION = HEP-TH/9712189;%%
\bibitem [{\citenamefont {Witten}(1996)}]{hep-th/9512219}%
  \BibitemOpen
  \bibfield  {author} {\bibinfo {author} {\bibfnamefont {Edward}\ \bibnamefont
  {Witten}},\ }\bibfield  {title} {\enquote {\bibinfo {title} {{Five-branes and
  M theory on an orbifold}},}\ }\href {\doibase 10.1016/0550-3213(96)00032-6}
  {\bibfield  {journal} {\bibinfo  {journal} {Nucl. Phys.}\ }\textbf {\bibinfo
  {volume} {B463}},\ \bibinfo {pages} {383--397} (\bibinfo {year} {1996})},\
  \bibinfo {note} {[,172(1995)]},\ \Eprint
  {http://arxiv.org/abs/hep-th/9512219} {arXiv:hep-th/9512219 [hep-th]}
  \BibitemShut {NoStop}%
%%CITATION = HEP-TH/9512219;%%
\bibitem [{\citenamefont {Uranga}(2002)}]{hep-th/0201221}%
  \BibitemOpen
  \bibfield  {author} {\bibinfo {author} {\bibfnamefont {Angel~M.}\
  \bibnamefont {Uranga}},\ }\bibfield  {title} {\enquote {\bibinfo {title}
  {{D-brane, fluxes and chirality}},}\ }\href {\doibase
  10.1088/1126-6708/2002/04/016} {\bibfield  {journal} {\bibinfo  {journal}
  {JHEP}\ }\textbf {\bibinfo {volume} {04}},\ \bibinfo {pages} {016} (\bibinfo
  {year} {2002})},\ \Eprint {http://arxiv.org/abs/hep-th/0201221}
  {arXiv:hep-th/0201221 [hep-th]} \BibitemShut {NoStop}%
%%CITATION = HEP-TH/0201221;%%
\bibitem [{\citenamefont {Romans}(1986)}]{Romans:1985tz}%
  \BibitemOpen
  \bibfield  {author} {\bibinfo {author} {\bibfnamefont {L.~J.}\ \bibnamefont
  {Romans}},\ }\bibfield  {title} {\enquote {\bibinfo {title} {{Massive N=2a
  Supergravity in Ten-Dimensions}},}\ }\href {\doibase
  10.1016/0370-2693(86)90375-8} {\bibfield  {journal} {\bibinfo  {journal}
  {Phys. Lett.}\ }\textbf {\bibinfo {volume} {B169}},\ \bibinfo {pages} {374}
  (\bibinfo {year} {1986})},\ \bibinfo {note} {[,374(1985)]}\BibitemShut
  {NoStop}%
%%CITATION = PHLTA,B169,374;%%
\bibitem [{\citenamefont {Bergshoeff}\ and\ \citenamefont
  {De~Roo}(1996)}]{hep-th/9603123}%
  \BibitemOpen
  \bibfield  {author} {\bibinfo {author} {\bibfnamefont {E.}~\bibnamefont
  {Bergshoeff}}\ and\ \bibinfo {author} {\bibfnamefont {M.}~\bibnamefont
  {De~Roo}},\ }\bibfield  {title} {\enquote {\bibinfo {title} {{D-branes and
  T-duality}},}\ }\href {\doibase 10.1016/0370-2693(96)00523-0} {\bibfield
  {journal} {\bibinfo  {journal} {Phys. Lett.}\ }\textbf {\bibinfo {volume}
  {B380}},\ \bibinfo {pages} {265--272} (\bibinfo {year} {1996})},\ \Eprint
  {http://arxiv.org/abs/hep-th/9603123} {arXiv:hep-th/9603123 [hep-th]}
  \BibitemShut {NoStop}%
%%CITATION = HEP-TH/9603123;%%
\bibitem [{\citenamefont {Green}\ \emph {et~al.}(1996)\citenamefont {Green},
  \citenamefont {Hull},\ and\ \citenamefont {Townsend}}]{hep-th/9604119}%
  \BibitemOpen
  \bibfield  {author} {\bibinfo {author} {\bibfnamefont {Michael~B.}\
  \bibnamefont {Green}}, \bibinfo {author} {\bibfnamefont {Christopher~M.}\
  \bibnamefont {Hull}}, \ and\ \bibinfo {author} {\bibfnamefont {Paul~K.}\
  \bibnamefont {Townsend}},\ }\bibfield  {title} {\enquote {\bibinfo {title}
  {{D-brane Wess-Zumino actions, T-duality and the cosmological constant}},}\
  }\href {\doibase 10.1016/0370-2693(96)00643-0} {\bibfield  {journal}
  {\bibinfo  {journal} {Phys. Lett.}\ }\textbf {\bibinfo {volume} {B382}},\
  \bibinfo {pages} {65--72} (\bibinfo {year} {1996})},\ \Eprint
  {http://arxiv.org/abs/hep-th/9604119} {arXiv:hep-th/9604119 [hep-th]}
  \BibitemShut {NoStop}%
%%CITATION = HEP-TH/9604119;%%
\bibitem [{\citenamefont {Frey}(2020)}]{freynew}%
  \BibitemOpen
  \bibfield  {author} {\bibinfo {author} {\bibfnamefont {Andrew~R.}\
  \bibnamefont {Frey}},\ }\bibfield  {title} {\enquote {\bibinfo {title}
  {{Dirac branes for Dirichlet branes: Brane degrees of freedom}},}\
  }\href@noop {} {\bibfield  {journal} {\bibinfo  {journal} {work in progress}\
  } (\bibinfo {year} {2020})}\BibitemShut {NoStop}%
\bibitem [{\citenamefont {Griffiths}\ and\ \citenamefont
  {Harris}(1994)}]{griffithsharris}%
  \BibitemOpen
  \bibfield  {author} {\bibinfo {author} {\bibfnamefont {Phillip}\ \bibnamefont
  {Griffiths}}\ and\ \bibinfo {author} {\bibfnamefont {Joseph}\ \bibnamefont
  {Harris}},\ }\href@noop {} {\emph {\bibinfo {title} {{Principles of algebraic
  geometry}}}},\ Wiley classics library ed.\ (\bibinfo  {publisher} {Wiley},\
  \bibinfo {year} {1994})\BibitemShut {NoStop}%
\bibitem [{\citenamefont {Cheung}\ and\ \citenamefont
  {Yin}(1998)}]{hep-th/9710206}%
  \BibitemOpen
  \bibfield  {author} {\bibinfo {author} {\bibfnamefont {Yeuk-Kwan~E.}\
  \bibnamefont {Cheung}}\ and\ \bibinfo {author} {\bibfnamefont {Zheng}\
  \bibnamefont {Yin}},\ }\bibfield  {title} {\enquote {\bibinfo {title}
  {{Anomalies, branes, and currents}},}\ }\href {\doibase
  10.1016/S0550-3213(98)00115-1} {\bibfield  {journal} {\bibinfo  {journal}
  {Nucl. Phys.}\ }\textbf {\bibinfo {volume} {B517}},\ \bibinfo {pages}
  {69--91} (\bibinfo {year} {1998})},\ \Eprint
  {http://arxiv.org/abs/hep-th/9710206} {arXiv:hep-th/9710206 [hep-th]}
  \BibitemShut {NoStop}%
%%CITATION = HEP-TH/9710206;%%
\bibitem [{\citenamefont {Freed}\ and\ \citenamefont
  {Witten}(1999)}]{hep-th/9907189}%
  \BibitemOpen
  \bibfield  {author} {\bibinfo {author} {\bibfnamefont {Daniel~S.}\
  \bibnamefont {Freed}}\ and\ \bibinfo {author} {\bibfnamefont {Edward}\
  \bibnamefont {Witten}},\ }\bibfield  {title} {\enquote {\bibinfo {title}
  {{Anomalies in string theory with D-branes}},}\ }\href@noop {} {\bibfield
  {journal} {\bibinfo  {journal} {Asian J. Math.}\ }\textbf {\bibinfo {volume}
  {3}},\ \bibinfo {pages} {819} (\bibinfo {year} {1999})},\ \Eprint
  {http://arxiv.org/abs/hep-th/9907189} {arXiv:hep-th/9907189 [hep-th]}
  \BibitemShut {NoStop}%
%%CITATION = HEP-TH/9907189;%%
\bibitem [{\citenamefont {Frey}\ and\ \citenamefont
  {Polchinski}(2002)}]{hep-th/0201029}%
  \BibitemOpen
  \bibfield  {author} {\bibinfo {author} {\bibfnamefont {Andrew~R.}\
  \bibnamefont {Frey}}\ and\ \bibinfo {author} {\bibfnamefont {Joseph}\
  \bibnamefont {Polchinski}},\ }\bibfield  {title} {\enquote {\bibinfo {title}
  {{N=3 warped compactifications}},}\ }\href {\doibase
  10.1103/PhysRevD.65.126009} {\bibfield  {journal} {\bibinfo  {journal} {Phys.
  Rev.}\ }\textbf {\bibinfo {volume} {D65}},\ \bibinfo {pages} {126009}
  (\bibinfo {year} {2002})},\ \Eprint {http://arxiv.org/abs/hep-th/0201029}
  {arXiv:hep-th/0201029 [hep-th]} \BibitemShut {NoStop}%
%%CITATION = HEP-TH/0201029;%%
\bibitem [{\citenamefont {Kashani-Poor}\ and\ \citenamefont
  {Tomasiello}(2005)}]{hep-th/0505208}%
  \BibitemOpen
  \bibfield  {author} {\bibinfo {author} {\bibfnamefont {Amir-Kian}\
  \bibnamefont {Kashani-Poor}}\ and\ \bibinfo {author} {\bibfnamefont
  {Alessandro}\ \bibnamefont {Tomasiello}},\ }\bibfield  {title} {\enquote
  {\bibinfo {title} {{A stringy test of flux-induced isometry gauging}},}\
  }\href {\doibase 10.1016/j.nuclphysb.2005.08.040} {\bibfield  {journal}
  {\bibinfo  {journal} {Nucl. Phys.}\ }\textbf {\bibinfo {volume} {B728}},\
  \bibinfo {pages} {135--147} (\bibinfo {year} {2005})},\ \Eprint
  {http://arxiv.org/abs/hep-th/0505208} {arXiv:hep-th/0505208 [hep-th]}
  \BibitemShut {NoStop}%
%%CITATION = HEP-TH/0505208;%%
\bibitem [{\citenamefont {Poisson}\ \emph {et~al.}(2011)\citenamefont
  {Poisson}, \citenamefont {Pound},\ and\ \citenamefont
  {Vega}}]{arXiv:1102.0529}%
  \BibitemOpen
  \bibfield  {author} {\bibinfo {author} {\bibfnamefont {Eric}\ \bibnamefont
  {Poisson}}, \bibinfo {author} {\bibfnamefont {Adam}\ \bibnamefont {Pound}}, \
  and\ \bibinfo {author} {\bibfnamefont {Ian}\ \bibnamefont {Vega}},\
  }\bibfield  {title} {\enquote {\bibinfo {title} {{The Motion of point
  particles in curved spacetime}},}\ }\href {\doibase 10.12942/lrr-2011-7}
  {\bibfield  {journal} {\bibinfo  {journal} {Living Rev. Rel.}\ }\textbf
  {\bibinfo {volume} {14}},\ \bibinfo {pages} {7} (\bibinfo {year} {2011})},\
  \Eprint {http://arxiv.org/abs/1102.0529} {arXiv:1102.0529 [gr-qc]}
  \BibitemShut {NoStop}%
%%CITATION = ARXIV:1102.0529;%%
\bibitem [{\citenamefont {Andriot}\ and\ \citenamefont
  {Blåbäck}(2017)}]{arXiv:1609.00385}%
  \BibitemOpen
  \bibfield  {author} {\bibinfo {author} {\bibfnamefont {David}\ \bibnamefont
  {Andriot}}\ and\ \bibinfo {author} {\bibfnamefont {Johan}\ \bibnamefont
  {Blåbäck}},\ }\bibfield  {title} {\enquote {\bibinfo {title} {{Refining the
  boundaries of the classical de Sitter landscape}},}\ }\href {\doibase
  10.1007/JHEP03(2017)102, 10.1007/JHEP03(2018)083} {\bibfield  {journal}
  {\bibinfo  {journal} {JHEP}\ }\textbf {\bibinfo {volume} {03}},\ \bibinfo
  {pages} {102} (\bibinfo {year} {2017})},\ \bibinfo {note} {[Erratum:
  JHEP03,083(2018)]},\ \Eprint {http://arxiv.org/abs/1609.00385}
  {arXiv:1609.00385 [hep-th]} \BibitemShut {NoStop}%
%%CITATION = ARXIV:1609.00385;%%
\bibitem [{\citenamefont {Green}\ \emph {et~al.}(1997)\citenamefont {Green},
  \citenamefont {Harvey},\ and\ \citenamefont {Moore}}]{hep-th/9605033}%
  \BibitemOpen
  \bibfield  {author} {\bibinfo {author} {\bibfnamefont {Michael~B.}\
  \bibnamefont {Green}}, \bibinfo {author} {\bibfnamefont {Jeffrey~A.}\
  \bibnamefont {Harvey}}, \ and\ \bibinfo {author} {\bibfnamefont {Gregory~W.}\
  \bibnamefont {Moore}},\ }\bibfield  {title} {\enquote {\bibinfo {title}
  {{I-brane inflow and anomalous couplings on D-branes}},}\ }\href {\doibase
  10.1088/0264-9381/14/1/008} {\bibfield  {journal} {\bibinfo  {journal}
  {Class. Quant. Grav.}\ }\textbf {\bibinfo {volume} {14}},\ \bibinfo {pages}
  {47--52} (\bibinfo {year} {1997})},\ \Eprint
  {http://arxiv.org/abs/hep-th/9605033} {arXiv:hep-th/9605033 [hep-th]}
  \BibitemShut {NoStop}%
%%CITATION = HEP-TH/9605033;%%
\bibitem [{\citenamefont {Johnson}(2005)}]{Johnson:2003gi}%
  \BibitemOpen
  \bibfield  {author} {\bibinfo {author} {\bibfnamefont {Clifford~V.}\
  \bibnamefont {Johnson}},\ }\href {\doibase 10.1017/CBO9780511606540} {\emph
  {\bibinfo {title} {{D-branes}}}},\ Cambridge Monographs on Mathematical
  Physics\ (\bibinfo  {publisher} {Cambridge University Press},\ \bibinfo
  {year} {2005})\BibitemShut {NoStop}%
%%CITATION = INSPIRE-612496;%%
\bibitem [{\citenamefont {Giddings}\ \emph {et~al.}(2002)\citenamefont
  {Giddings}, \citenamefont {Kachru},\ and\ \citenamefont
  {Polchinski}}]{hep-th/0105097}%
  \BibitemOpen
  \bibfield  {author} {\bibinfo {author} {\bibfnamefont {Steven~B.}\
  \bibnamefont {Giddings}}, \bibinfo {author} {\bibfnamefont {Shamit}\
  \bibnamefont {Kachru}}, \ and\ \bibinfo {author} {\bibfnamefont {Joseph}\
  \bibnamefont {Polchinski}},\ }\bibfield  {title} {\enquote {\bibinfo {title}
  {{Hierarchies from fluxes in string compactifications}},}\ }\href {\doibase
  10.1103/PhysRevD.66.106006} {\bibfield  {journal} {\bibinfo  {journal} {Phys.
  Rev.}\ }\textbf {\bibinfo {volume} {D66}},\ \bibinfo {pages} {106006}
  (\bibinfo {year} {2002})},\ \Eprint {http://arxiv.org/abs/hep-th/0105097}
  {arXiv:hep-th/0105097 [hep-th]} \BibitemShut {NoStop}%
%%CITATION = HEP-TH/0105097;%%
\bibitem [{\citenamefont {Pasti}\ \emph
  {et~al.}(1997{\natexlab{a}})\citenamefont {Pasti}, \citenamefont {Sorokin},\
  and\ \citenamefont {Tonin}}]{hep-th/9611100}%
  \BibitemOpen
  \bibfield  {author} {\bibinfo {author} {\bibfnamefont {Paolo}\ \bibnamefont
  {Pasti}}, \bibinfo {author} {\bibfnamefont {Dmitri~P.}\ \bibnamefont
  {Sorokin}}, \ and\ \bibinfo {author} {\bibfnamefont {Mario}\ \bibnamefont
  {Tonin}},\ }\bibfield  {title} {\enquote {\bibinfo {title}
  {{Lorentz-invariant actions for chiral $p$-forms}},}\ }\href {\doibase
  10.1103/PhysRevD.55.6292} {\bibfield  {journal} {\bibinfo  {journal} {Phys.
  Rev.}\ }\textbf {\bibinfo {volume} {D55}},\ \bibinfo {pages} {6292--6298}
  (\bibinfo {year} {1997}{\natexlab{a}})},\ \Eprint
  {http://arxiv.org/abs/hep-th/9611100} {arXiv:hep-th/9611100 [hep-th]}
  \BibitemShut {NoStop}%
%%CITATION = HEP-TH/9611100;%%
\bibitem [{\citenamefont {Pasti}\ \emph
  {et~al.}(1997{\natexlab{b}})\citenamefont {Pasti}, \citenamefont {Sorokin},\
  and\ \citenamefont {Tonin}}]{hep-th/9701037}%
  \BibitemOpen
  \bibfield  {author} {\bibinfo {author} {\bibfnamefont {Paolo}\ \bibnamefont
  {Pasti}}, \bibinfo {author} {\bibfnamefont {Dmitri~P.}\ \bibnamefont
  {Sorokin}}, \ and\ \bibinfo {author} {\bibfnamefont {Mario}\ \bibnamefont
  {Tonin}},\ }\bibfield  {title} {\enquote {\bibinfo {title} {{Covariant action
  for a D = 11 five-brane with the chiral field}},}\ }\href {\doibase
  10.1016/S0370-2693(97)00188-3} {\bibfield  {journal} {\bibinfo  {journal}
  {Phys. Lett.}\ }\textbf {\bibinfo {volume} {B398}},\ \bibinfo {pages}
  {41--46} (\bibinfo {year} {1997}{\natexlab{b}})},\ \Eprint
  {http://arxiv.org/abs/hep-th/9701037} {arXiv:hep-th/9701037 [hep-th]}
  \BibitemShut {NoStop}%
%%CITATION = HEP-TH/9701037;%%
\bibitem [{\citenamefont {Sen}(2016)}]{arXiv:1511.08220}%
  \BibitemOpen
  \bibfield  {author} {\bibinfo {author} {\bibfnamefont {Ashoke}\ \bibnamefont
  {Sen}},\ }\bibfield  {title} {\enquote {\bibinfo {title} {{Covariant Action
  for Type IIB Supergravity}},}\ }\href {\doibase 10.1007/JHEP07(2016)017}
  {\bibfield  {journal} {\bibinfo  {journal} {JHEP}\ }\textbf {\bibinfo
  {volume} {07}},\ \bibinfo {pages} {017} (\bibinfo {year} {2016})},\ \Eprint
  {http://arxiv.org/abs/1511.08220} {arXiv:1511.08220 [hep-th]} \BibitemShut
  {NoStop}%
%%CITATION = ARXIV:1511.08220;%%
\bibitem [{\citenamefont {Sen}(2020)}]{arXiv:1903.12196}%
  \BibitemOpen
  \bibfield  {author} {\bibinfo {author} {\bibfnamefont {Ashoke}\ \bibnamefont
  {Sen}},\ }\bibfield  {title} {\enquote {\bibinfo {title} {{Self-dual forms:
  Action, Hamiltonian and Compactification}},}\ }\href {\doibase
  10.1088/1751-8121/ab5423} {\bibfield  {journal} {\bibinfo  {journal} {J.
  Phys. A}\ }\textbf {\bibinfo {volume} {53}},\ \bibinfo {pages} {084002}
  (\bibinfo {year} {2020})},\ \Eprint {http://arxiv.org/abs/1903.12196}
  {arXiv:1903.12196 [hep-th]} \BibitemShut {NoStop}%
\bibitem [{\citenamefont {Frey}\ \emph {et~al.}(2009)\citenamefont {Frey},
  \citenamefont {Torroba}, \citenamefont {Underwood},\ and\ \citenamefont
  {Douglas}}]{arXiv:0810.5768}%
  \BibitemOpen
  \bibfield  {author} {\bibinfo {author} {\bibfnamefont {Andrew~R.}\
  \bibnamefont {Frey}}, \bibinfo {author} {\bibfnamefont {Gonzalo}\
  \bibnamefont {Torroba}}, \bibinfo {author} {\bibfnamefont {Bret}\
  \bibnamefont {Underwood}}, \ and\ \bibinfo {author} {\bibfnamefont
  {Michael~R.}\ \bibnamefont {Douglas}},\ }\bibfield  {title} {\enquote
  {\bibinfo {title} {{The Universal K\"ahler Modulus in Warped
  Compactifications}},}\ }\href {\doibase 10.1088/1126-6708/2009/01/036}
  {\bibfield  {journal} {\bibinfo  {journal} {JHEP}\ }\textbf {\bibinfo
  {volume} {01}},\ \bibinfo {pages} {036} (\bibinfo {year} {2009})},\ \Eprint
  {http://arxiv.org/abs/0810.5768} {arXiv:0810.5768 [hep-th]} \BibitemShut
  {NoStop}%
%%CITATION = ARXIV:0810.5768;%%
\bibitem [{\citenamefont {Frey}\ and\ \citenamefont
  {Roberts}(2013)}]{arXiv:1308.0323}%
  \BibitemOpen
  \bibfield  {author} {\bibinfo {author} {\bibfnamefont {Andrew~R.}\
  \bibnamefont {Frey}}\ and\ \bibinfo {author} {\bibfnamefont {James}\
  \bibnamefont {Roberts}},\ }\bibfield  {title} {\enquote {\bibinfo {title}
  {{The Dimensional Reduction and Kähler Metric of Forms In Flux and
  Warping}},}\ }\href {\doibase 10.1007/JHEP10(2013)021} {\bibfield  {journal}
  {\bibinfo  {journal} {JHEP}\ }\textbf {\bibinfo {volume} {10}},\ \bibinfo
  {pages} {021} (\bibinfo {year} {2013})},\ \Eprint
  {http://arxiv.org/abs/1308.0323} {arXiv:1308.0323 [hep-th]} \BibitemShut
  {NoStop}%
%%CITATION = ARXIV:1308.0323;%%
\bibitem [{\citenamefont {Cownden}\ \emph {et~al.}(2016)\citenamefont
  {Cownden}, \citenamefont {Frey}, \citenamefont {Marsh},\ and\ \citenamefont
  {Underwood}}]{arXiv:1609.05904}%
  \BibitemOpen
  \bibfield  {author} {\bibinfo {author} {\bibfnamefont {Brad}\ \bibnamefont
  {Cownden}}, \bibinfo {author} {\bibfnamefont {Andrew~R.}\ \bibnamefont
  {Frey}}, \bibinfo {author} {\bibfnamefont {M.~C.~David}\ \bibnamefont
  {Marsh}}, \ and\ \bibinfo {author} {\bibfnamefont {Bret}\ \bibnamefont
  {Underwood}},\ }\bibfield  {title} {\enquote {\bibinfo {title} {{Dimensional
  Reduction for D3-brane Moduli}},}\ }\href {\doibase 10.1007/JHEP12(2016)139}
  {\bibfield  {journal} {\bibinfo  {journal} {JHEP}\ }\textbf {\bibinfo
  {volume} {12}},\ \bibinfo {pages} {139} (\bibinfo {year} {2016})},\ \Eprint
  {http://arxiv.org/abs/1609.05904} {arXiv:1609.05904 [hep-th]} \BibitemShut
  {NoStop}%
%%CITATION = ARXIV:1609.05904;%%
\bibitem [{\citenamefont {Dall'Agata}\ \emph {et~al.}(1997)\citenamefont
  {Dall'Agata}, \citenamefont {Lechner},\ and\ \citenamefont
  {Sorokin}}]{DallAgata:1997gnw}%
  \BibitemOpen
  \bibfield  {author} {\bibinfo {author} {\bibfnamefont {Gianguido}\
  \bibnamefont {Dall'Agata}}, \bibinfo {author} {\bibfnamefont {Kurt}\
  \bibnamefont {Lechner}}, \ and\ \bibinfo {author} {\bibfnamefont {Dmitri~P.}\
  \bibnamefont {Sorokin}},\ }\bibfield  {title} {\enquote {\bibinfo {title}
  {{Covariant actions for the bosonic sector of d = 10 IIB supergravity}},}\
  }\href {\doibase 10.1088/0264-9381/14/12/003} {\bibfield  {journal} {\bibinfo
   {journal} {Class. Quant. Grav.}\ }\textbf {\bibinfo {volume} {14}},\
  \bibinfo {pages} {L195--L198} (\bibinfo {year} {1997})},\ \Eprint
  {http://arxiv.org/abs/hep-th/9707044} {arXiv:hep-th/9707044 [hep-th]}
  \BibitemShut {NoStop}%
%%CITATION = HEP-TH/9707044;%%
\bibitem [{\citenamefont {Polchinski}(1995)}]{hep-th/9510017}%
  \BibitemOpen
  \bibfield  {author} {\bibinfo {author} {\bibfnamefont {Joseph}\ \bibnamefont
  {Polchinski}},\ }\bibfield  {title} {\enquote {\bibinfo {title} {{Dirichlet
  Branes and Ramond-Ramond charges}},}\ }\href {\doibase
  10.1103/PhysRevLett.75.4724} {\bibfield  {journal} {\bibinfo  {journal}
  {Phys. Rev. Lett.}\ }\textbf {\bibinfo {volume} {75}},\ \bibinfo {pages}
  {4724--4727} (\bibinfo {year} {1995})},\ \Eprint
  {http://arxiv.org/abs/hep-th/9510017} {arXiv:hep-th/9510017 [hep-th]}
  \BibitemShut {NoStop}%
%%CITATION = HEP-TH/9510017;%%
\bibitem [{\citenamefont {Polchinski}\ and\ \citenamefont
  {Witten}(1996)}]{hep-th/9510169}%
  \BibitemOpen
  \bibfield  {author} {\bibinfo {author} {\bibfnamefont {Joseph}\ \bibnamefont
  {Polchinski}}\ and\ \bibinfo {author} {\bibfnamefont {Edward}\ \bibnamefont
  {Witten}},\ }\bibfield  {title} {\enquote {\bibinfo {title} {{Evidence for
  heterotic - type I string duality}},}\ }\href {\doibase
  10.1016/0550-3213(95)00614-1} {\bibfield  {journal} {\bibinfo  {journal}
  {Nucl. Phys.}\ }\textbf {\bibinfo {volume} {B460}},\ \bibinfo {pages}
  {525--540} (\bibinfo {year} {1996})},\ \Eprint
  {http://arxiv.org/abs/hep-th/9510169} {arXiv:hep-th/9510169 [hep-th]}
  \BibitemShut {NoStop}%
%%CITATION = HEP-TH/9510169;%%
\bibitem [{\citenamefont {Polchinski}\ and\ \citenamefont
  {Strominger}(1996)}]{hep-th/9510227}%
  \BibitemOpen
  \bibfield  {author} {\bibinfo {author} {\bibfnamefont {Joseph}\ \bibnamefont
  {Polchinski}}\ and\ \bibinfo {author} {\bibfnamefont {Andrew}\ \bibnamefont
  {Strominger}},\ }\bibfield  {title} {\enquote {\bibinfo {title} {{New vacua
  for type II string theory}},}\ }\href {\doibase
  10.1016/S0370-2693(96)01219-1} {\bibfield  {journal} {\bibinfo  {journal}
  {Phys. Lett.}\ }\textbf {\bibinfo {volume} {B388}},\ \bibinfo {pages}
  {736--742} (\bibinfo {year} {1996})},\ \Eprint
  {http://arxiv.org/abs/hep-th/9510227} {arXiv:hep-th/9510227 [hep-th]}
  \BibitemShut {NoStop}%
%%CITATION = HEP-TH/9510227;%%
\bibitem [{\citenamefont {Myers}(1999)}]{Myers:1999ps}%
  \BibitemOpen
  \bibfield  {author} {\bibinfo {author} {\bibfnamefont {Robert~C.}\
  \bibnamefont {Myers}},\ }\bibfield  {title} {\enquote {\bibinfo {title}
  {{Dielectric branes}},}\ }\href {\doibase 10.1088/1126-6708/1999/12/022}
  {\bibfield  {journal} {\bibinfo  {journal} {JHEP}\ }\textbf {\bibinfo
  {volume} {12}},\ \bibinfo {pages} {022} (\bibinfo {year} {1999})},\ \Eprint
  {http://arxiv.org/abs/hep-th/9910053} {arXiv:hep-th/9910053 [hep-th]}
  \BibitemShut {NoStop}%
%%CITATION = HEP-TH/9910053;%%
\bibitem [{\citenamefont {Polchinski}(1998)}]{Polchinski:1998rr}%
  \BibitemOpen
  \bibfield  {author} {\bibinfo {author} {\bibfnamefont {J.}~\bibnamefont
  {Polchinski}},\ }\href {\doibase 10.1017/CBO9780511618123} {\emph {\bibinfo
  {title} {{String theory. Vol. 2: Superstring theory and beyond}}}},\
  Cambridge Monographs on Mathematical Physics\ (\bibinfo  {publisher}
  {Cambridge University Press},\ \bibinfo {year} {1998})\BibitemShut {NoStop}%
%%CITATION = INSPIRE-487241;%%
\end{thebibliography}%

\end{document}